\documentclass[onecolumn,authoryear]{els-mrw}

\usepackage{amsmath,amssymb,amsfonts,amsthm,makeidx,graphicx}
\usepackage{txfonts}
\usepackage{helvet}
\usepackage{graphicx}
\usepackage{color}

\begin{document}

\chapter{Rings around giant planets and smaller bodies}\label{chap1}

\author{Keiji Ohtsuki}%

\address[0]{\orgname{Kobe University}, \orgdiv{Department of Planetology}, \orgaddress{1-1 Rokkodai, Nada, Kobe 657-8501, Japan}}

\maketitle

\begin{abstract}[Abstract]
All the four giant planets in our Solar System have rings, but their characteristics are very different.
The rings consist of a number of small particles, although individual particles have not been directly imaged.
Near the central planet, colliding particles bounce off each other in low-velocity impacts but cannot gravitationally merge due to the effect of the tidal force, resulting in the formation of rings, whereas in more distant regions particles can gravitationally accrete to form satellites.
Rings exhibit various types of fine structure, and the mutual gravitational forces between particles and the gravity from satellites play an important role in rings of macroscopic particles, while non-gravitational forces are important for dusty rings.
There are several theories about the origin of rings, and formation mechanisms are likely to be different among different ring systems.
The rings of small Solar System bodies were discovered through observations of occultations of stars by these bodies.
It is natural to expect that some exoplanets should also have rings, but their detection remains challenging.
Future discovery of more ring-moon systems around small bodies and exoplanets will provide clues to understanding the formation and evolution of the central bodies that host them.
\end{abstract}

\vspace{4mm}
\noindent
{\fontsize{10}{12}\selectfont {\bf \textsf{Keywords:}}} Planetary rings, Jupiter, Saturn, Uranus, Neptune, Small Solar System bodies, Jovian satellites, Saturnian satellites, Uranian satellites, Neptunian satellites

\vspace{4mm}
\noindent
\hrulefill

\noindent\textbf{Glossary} \hfill 1 \\
\noindent\textbf{1. Introcution} \hfill 2 \\
\noindent\textbf{2. Preliminaries} \hfill 2 \\
\noindent\textbf{3. Overview of each ring system} \hfill 5 \\
\noindent\hspace*{4mm}{3.1 The rings of Saturn} \hfill 5 \\
\noindent\hspace*{4mm}{3.2 The rings of Uranus} \hfill 8 \\
\noindent\hspace*{4mm}{3.3 The rings of Jupiter} \hfill 10 \\
\noindent\hspace*{4mm}{3.4 The rings of Neptune} \hfill 10 \\
\noindent\hspace*{4mm}{3.5 Rings of small solar system bodies} \hfill 12 \\
\noindent\textbf{4. Origins of ring systems} \hfill 13 \\
\noindent\textbf{5. Concluding remarks} \hfill 14 \\
\noindent\textbf{Acknowledgments} \hfill 15 \\
\noindent\textbf{References} \hfill 15 \\

\vspace{-6mm}
\noindent
\hrulefill

\begin{glossary}[Glossary]

\term{Coefficient of restitution:}
A dimensionless parameter defined as the ratio of the relative velocity of separation after a collision of two bodies to the relative velocity of approach before the collision. More energy is dissipated in the collision with a smaller value of the coefficient of restitution.

\term{Dynamical optical depth:}
A dimensionless quantity used in numerical simulation of planetary rings and is defined by the sum of the cross-section areas of particles used in the simulation divided by the area of the simulation region. 

\term{Epicyclic frequency:}
The frequency at which a particle on an eccentric orbit about a central body will oscillate in the radial direction. It corresponds to the orbital angular velocity in the case of the Keplerian rotation around a spherically symmetric central body.

\term{Self-gravity wakes:} 
Non-axisymmetric structure formed in sufficiently dense planetary rings due to the competition between collective gravity and differential rotation. Also called  gravitational wakes.

\term{Optical depth:}
A dimensionless quantity that determines the reduction in light intensity due to absorption and scattering. 
When light from a source with intensity $I_0$ passes through a material and the intensity becomes $I$, the optical depth is defined using the transmittance $T=I/I_0$ as $\tau = -\ln T$. From this, $I = I_0 \exp^{-\tau}$.
A greater optical depth corresponds to a greater reduction in light intensity. 
Also called optical thickness. 

\term{Orbital evolution due to tidal force:}
The gravitational force from a satellite causes deformation of a planet toward the satellite, which is called the tidal bulge. Because the planet is not perfect fluid and its deformation requires a finite response time, a lag can arise between the direction of the satellite and that of the tidal bulge on the planet. If the planet is spinning faster (slower) than the orbital angular velocity of the satellite, the gravitational attraction between the tidal bulge and the satellite slows down (speeds up) the planet's spin and expands (shrinks) the satellite's orbit, resulting in orbital evolution of the satellite due to the tidal force.

\term{Poynting-Robertson drag:}
A dust particle orbiting a star absorbs radiation from the star and re-radiates the energy isotropically in its own frame. From the inertial frame of the star, this means that the particle radiates preferentially in the direction of its orbital motion, which results in the loss of the particle's energy and angular momentum, leading to orbital decay toward the star. This effect is called Poynting-Robertson drag. 

\term{Stellar occultation:} 
An event where the light from a star is blocked by an object (such as a planet, moon, ring, or asteroid) between the star and an observer.
Observation of such an event can be used for a high-accuracy determination of the size and shape of the intervening body, an investigation of its atmosphere, and the detection of ring systems around the body.

\term{Tidal disruption:} 
When a small body passes very close to a large object such as a planet and the the tidal forces exerted on the small body by the large object exeeds the small body's internal strength and self-gravity, the small body will be pulled apart. This phenomenon is called tidal disruption.

\term{Tidal force:} 
The gravitational forces on an object caused by an external object vary on different parts of the object. The tidal force is created by these differential gravitational forces.

\term{Toomre $Q$ parameter:} 
A dimensionless quantity that determines axisymmetric stability of self-gravitating rotating disks. In the case of particle disks, it is defined as $Q=c_r \kappa/(3.36 G\Sigma)$, where $c_r$ is the radial component of the particles' velocity dispersion, $\kappa$ is the epicyclic frequency, $G$ is the gravitational constant, and $\Sigma$ is the surface density of the disk. 
The disk is stable if $Q>1$, which is known as Toomre's stability criterion.
However, whenever $Q \simeq 1-2$, the ring is susceptible to development of non-axisymmetric self-gravitational wakes.

\term{Velocity dispersion:}
The velocity dispersion is the statistical dispersion of velocities about the mean velocity. In the case of particles in planetary rings, it describes the degree of the deviation from locally non-inclined circular orbits. High velocity dispersion corresponds to highly eccentric and/or inclined orbits.

\end{glossary}

\section{Introduction}\label{chap_ring:sec1}

Until the mid-1970s, Saturn was the only planet with known rings, having been first seen by humans in the 1600s with the aid of newly invented telescopes.
In 1977, the rings of Uranus were unexpectedly discovered through observations from Earth, from an airplane equipped with a telescope \citep{Ell1977}, and later, the Voyager 1 spacecraft discovered a ring around Jupiter \citep{Owe1979}. 
Then, ten years later, Voyager 2 confirmed the rings around Neptune \citep{Smi1989}, although the presence of a partial ring was previously reported based on ground-based stellar occultation observations \citep{Hub1986}. 
This resulted in all the giant planets in the Solar System having rings, suggesting that rings are a ubiquitous feature of giant planet systems.

However, rings are not exclusive to giant planets. 
In 2013, rings were discovered around a small Solar System body, Chariklo \citep{Bra2014}, which is about 250 km in diameter with an orbit between Saturn and Uranus, and rings have also been confirmed around a few more small bodies.
This discovery significantly enhanced the importance of the study of ring systems in planetary science.

As we will show below, ring systems discovered so far in our Solar System are diverse in terms of shape, brightness, and number, and it is thought that the formation process of these systems is also diverse.
This means that ring systems can provide important clues for exploring the formation processes of the parent objects, whether they be planets or small bodies, and also the evolutionary processes that these bodies have undergone since their formation.

This chapter provides an overview of rings around giant planets and small Solar System bodies.
First, Section 2 briefly explains the basic physical processes and terminology related to rings.
Section 3 describes the characteristics of each ring system, from Saturn's rings to rings of small bodies, in the order of discovery.
Section 4 briefly describes the origin of rings, especially focusing on Saturn's rings.
Concluding remarks of this chapter are given in Section 5.

\section{Preliminaries}\label{chap_ring:sec2}

Rings that we will discuss in this chapter consist of numerous particles that revolve around a central body (planet, asteroid).
Individual particles are too small to be directly imaged. 
Even the images taken by the Cassini spacecraft, which orbited Saturn from 2004 to 2017 and performed various detailed observations of its rings, did not have enough resolution to identify individual particles.

For ring systems of the giant planets, information on particle size and particle composition have been obtained indirectly from remote sensing observations.
The size of the particles varies depending on the ring system, ranging from microns to the order of 10 m.
The composition of the particles also varies between ring systems.
Observations show that in Saturn's rings particles are more than 90-95\% water ice, while those in the rings of Uranus and Neptune seem to be made of dark materials, either carbonaceous chondrite-like materials or organic matter darkened by radiation. 
This information provides important clues to the origin of these rings.

The optical depth (also called optical thickness) $\tau$ of a ring is an important observable quantity that can be used to infer the amount of material in the ring. This dimensionless quantity determines the reduction in light intensity due to absorption and scattering. 
As a similar parameter, dynamical optical depth $\tau_\mathrm{dyn}$ is used in theoretical models of rings and is defined by the sum of the cross-section areas of particles used in the simulation divided by the area of the simulation region.
$\tau \simeq \tau_\mathrm{dyn}$ in dilute rings of randomly distributed particles, but they can be different in dense rings with non-uniform particle distribution  (Section \ref{chap_ring:subsec31}).

\begin{figure}[htb]
\centering
\includegraphics[bb=68.996100 79.238300 569.386000 734.308000, width=0.7\textwidth]{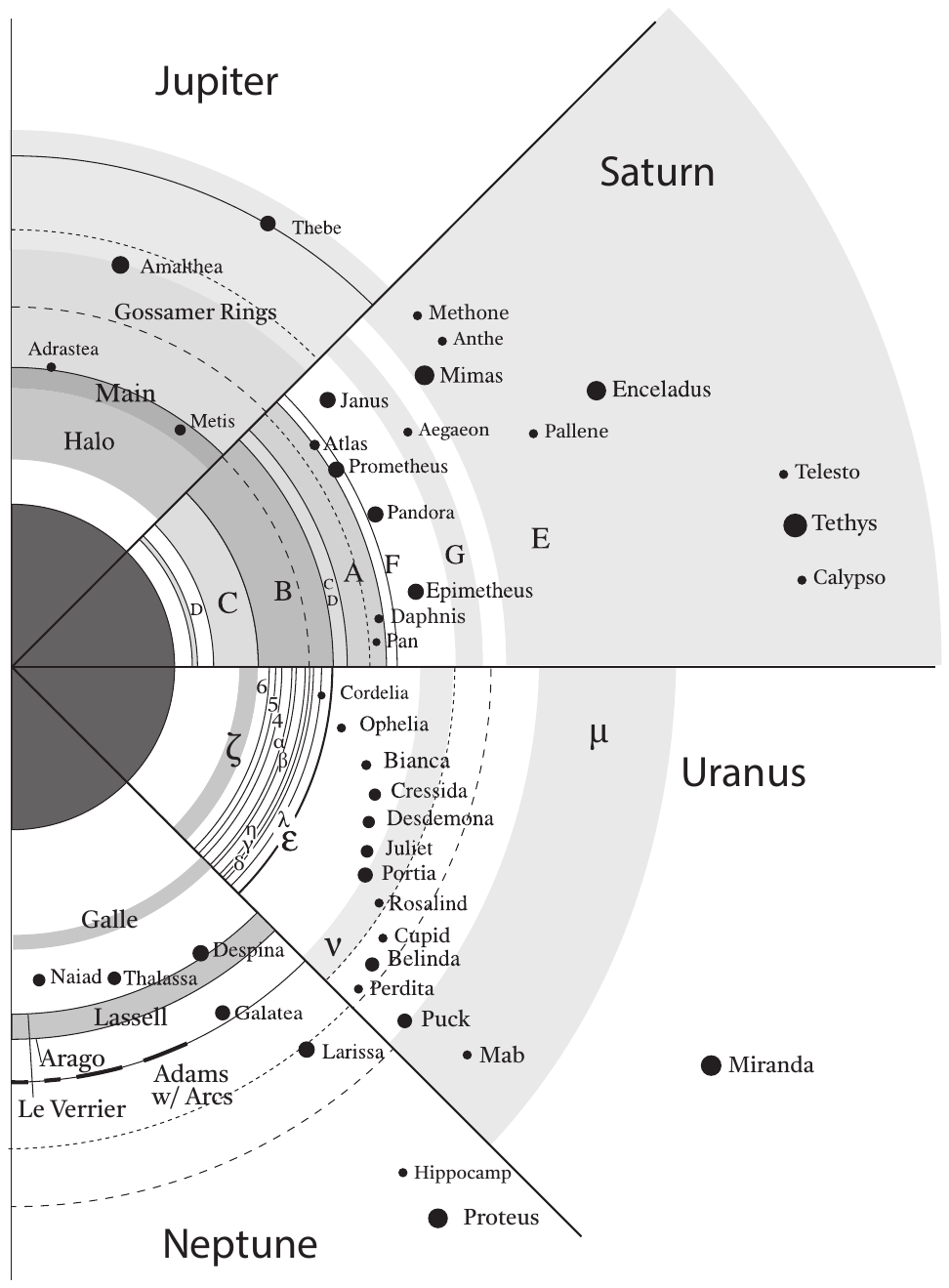}
\caption{A schematic of ring-moon systems of the four giant planets scaled to a common planetary radius. The planet is the solid central circle, ring regions are shaded, and nearby satellites are also shown. Dotted lines indicate the Roche limit for a satellite density of 1 g cm${}^{-3}$.
Dashed lines represent the position of synchronous orbit where an object's orbital period matches the rotation period of the planet.
``CD'' between Saturn's A and B rings refers to the Cassini Division.
Adapted from \citet{Bur2001}, ``Dusty rings and circumplanetary dust: Observations and simple physics.'' In: {Gr{\"u}n}, E. and {Gustafson}, B.~A.~S. and {Dermott}, S. and {Fechtig}, H. (Eds.), {\it Interplanetary Dust}, pp.641, Elsevier. Figure courtesy of Judith K. Burns and Douglas Hamilton.
Partly modified with assistance from Takaaki Takeda.
}
\label{fig:section2_f1}
\end{figure}

Saturn's main rings (the A, B, and C rings; see Figures \ref{fig:section2_f1} and \ref{fig:saturn1}) as well as the dense narrow rings of Uranus are mainly composed of macroscopic particles larger than cm in size.
In such rings, particles undergo repeated mutual collisions while orbiting around the central body.  
For example, the orbital speed of particles in the middle of Saturn's main rings is about 20 km s${}^{-1}$, but because the particles are moving almost in the same direction at almost the same speed, their relative velocity, or impact velocity, is rather slow, typically on the order of mm s${}^{-1}$ or even smaller \citep{Sal1995,Oht2020}.

Usually the orbits of ring particles are almost perfectly circular and coplanar, because inelastic collisions between the particles damp out the velocity deviations from the mean flow, which is the circular and coplanar motion.
The eccentricity $e$ of an orbit represents the degree of deviation from the circular orbit, and the inclination $i$ is the angle between the orbital plane and the reference plane (for example, the equatorial plane of the central object).
The velocity dispersion of particles is on the order of 
$\left(\langle e^2 \rangle+ \langle i^2 \rangle \right)^{1/2}v_\mathrm{K}$, 
where $v_\mathrm{K}$ is the velocity of a particle on the circular orbit at the given radial distance from the central body,
and $\langle e^2 \rangle$ and $\langle i^2 \rangle$ are the mean squares of particles' orbital eccentricities and inclinations, respectively.
In the case of collision between non-gravitating particles of equal radius $R$ initially on circular orbits, the impact velocity is determined by the Kepler shear due to the differential rotation and is on the order of $R\Omega$, 
where $\Omega$ is the orbital angular frequency,
so the collision causes a velocity dispersion on the order of $R\Omega$.
Particles' velocity dispersion increases because the energy of orbital motion is converted into energy of random motion through collisions, but the energy is dissipated due to inelastic collisions.
If the coefficient of restitution is sufficiently small, the balance between the two will result in an equilibrium state where 
$\langle e^2 \rangle$ and $\langle i^2 \rangle$ 
can be kept small, so the orbit will be approximately circular and the ring will have a thin disk structure.
On the other hand, if the coefficient of restitution is not small enough, energy dissipation due to inelastic collisions is not sufficient, and 
$\langle e^2 \rangle$ and $\langle i^2 \rangle$ 
will keep increasing.
Laboratory impact experiments show that the coefficient of restitution depends on impact velocity \citep{Col2018}.

The collision frequency depends on the particle spatial number density.
In dilute rings ($\tau_\mathrm{dyn} \ll 1$) consisting of equal-sized particles with radius $R$ with large velocity dispersion, the collision frequency is on the order of $\tau_\mathrm{dyn} \Omega$, but in dense rings the collision frequency can be much larger \citep{Sch2009}.
Furthermore, particles exert mutual gravitational forces on each other.
The importance of self-gravity in the ring depends on several factors, such as the surface density, distance from the central object, and particles' velocity dispersion, which depends on their coefficient of restitution \citep{Sal1995}.

When colliding ring particles gravitationally accrete and grow in mass, satellites can be formed, but usually this does not take place at the location where we find rings.
In the case of rings consisting of macroscopic particles, coalescence into a satellite is prevented by the tidal force from the central body.
When one particle is orbiting the central body, the region around the particle where the gravitational force from the particle is dominant over the tidal force from the central object is called the Hill sphere.
In fact, the shape of the Hill sphere is not exactly spherical but is lemon-shaped, stretched in the radial direction.
The size of the semi-axis of the Hill sphere in the radial direction is called Hill radius.
The Hill radius of a particle with mass $m$ orbiting at a radial distance $a$ about a central body with mass $M_\mathrm{c}$  is given as

\begin{equation}
{\displaystyle
R_\mathrm{H}=a\left( \frac{m}{3M_\mathrm{c}}\right)^{1/3}.
}
\end{equation}

\noindent
The size of 
the semi-axis of
the Hill sphere in the direction of orbital motion is $(2/3) R_\mathrm{H}$, while it is $(3^{2/3}-3^{1/3})R_\mathrm{ H} \simeq 0.638 R_\mathrm{H}$ in the vertical direction \citep{Yas2014}.
Therefore, the entire surface of a spherical particle will fall inside the Hill sphere when $R/ R_\mathrm{H} \lesssim 0.6$, and if there is sufficient energy dissipation in the collision, the colliding particles will coalesce due to mutual gravity. 
Gravitational accretion does not occur even when $R/ R_\mathrm{H} < 0.6$ if the coefficient of restitution and/or impact velocity is too large.
However, in the case of $R/ R_\mathrm{H} \gtrsim 0.7$, gravitational accretion does not take place because colliding particles rebound before entering the Hill sphere \citep{Oht1993}.
The effect of mutual gravity between colliding particles can be ignored when $ R_\mathrm{H}/R \ll 1$, while it can be important when $ R_\mathrm{H}/R \gtrsim 1$.

The bulk density of a body that entirely fills its Hill sphere defines a critical density $\rho_\mathrm{Roche}$ at its radial distance $a$ from the central body \citep{Por2007}

\begin{equation}
{\displaystyle
\rho_\mathrm{Roche} = \frac{3M_\mathrm{c}}{\gamma a^3},
}
\end{equation}

\noindent
where $\gamma$ is a dimensionless shape parameter so that $\gamma R_\mathrm{sat}^3$ is the volume of a satellite with $ R_\mathrm{sat}$ being its long semi-axis; $\gamma =4\pi/3$ for a spherical body and $\gamma \simeq 1.509$ for a body with a shape of the Hill sphere \citep{Lei2012}.
An aggregate with a bulk density $\rho_\mathrm{bulk}$ placed at a radial distance $a$ from the central body can be gravitationally held together when $\rho_\mathrm{bulk} > \rho_\mathrm{Roche} $.
This gives a critical radial distance from the central body beyond which the aggregate can be gravitationally held together:

\begin{equation}
 {\displaystyle
a_\mathrm{crit} =
\left( \frac{4\pi \rho_\mathrm{c}}{\gamma \rho_\mathrm{bulk}} \right)^{1/3} R_\mathrm{c},
}
\end{equation}

\noindent
where $ R_\mathrm{c}$ and $\rho_\mathrm{c}$ denote the radius and internal density of the central body.
Substituting $\gamma = 1.509$, we have $a_\mathrm{crit} \simeq 2.03 (\rho_\mathrm{c}/\rho_\mathrm{bulk})^{1/3} R_\mathrm{c}$.
This is similar to the so-called Roche limit for a self-gravitating fluid body with density $\rho$, which is given as \citep[e.g.,][]{Cha2018}

\begin{equation}
{\displaystyle
a_\mathrm{Roche} = 2.45 \left(\frac{\rho_\mathrm{c}}{\rho}\right)^{1/3}R_\mathrm{c}.
}
\label{eq:roche}
\end{equation}

Figure \ref{fig:section2_f1} shows a schematic of rings and nearby moons of the four giant planets scaled to a common planetary radius.
For each ring system, the Roche limit for a satellite density of 1 g cm${}^{-3}$ is shown by the dotted line.
If the density of the particles is lower than this, $a_\mathrm{Roche}$ will be larger,  while it will be smaller for a higher density.
For example, in the case of Saturn, the outer edge of the main rings is located slightly outside the Roche limit shown by the dotted line, suggesting that the rings are likely composed of porous particles with a density smaller than 1 g cm${}^{-3}$.
We can see a tendency that rings are located close to the planet and moons orbit at larger distances, which seems to be roughly explained by the Roche limit.
On the other hand, there are cases where moons are located inside the Roche limit and moons and rings coexist.
Such complicated distributions likely reflect not only differences in material density but also their evolutionary processes.
Interestingly, it can be shown from Eq.(\ref{eq:roche}) that the orbital angular velocity, or the mean motion, of particles near the Roche limit is approximately given as 
$\sqrt{GM_\mathrm{c}/a_\mathrm{Roche}^3} \sim 
\sqrt{GM_\mathrm{c}/a_\mathrm{crit}^3} = \sqrt{\gamma G \rho_\mathrm{bulk}/3}$, depending only on the material density of the particles if $\gamma$ is constant \citep{Sic2018}.
Thus, the magnitude of the minimum velocity dispersion ($\sim R\Omega$), which is roughly equal to the minimum impact velocity, of ring particles 
held together by weak gravity and 
located near the Roche limit is approximately determined only by the size and material density of the particles regardless of ring systems.
Figure \ref{fig:section2_f1} also shows the position of synchronous orbit where an object's orbital period matches the rotation period of the planet.
Owing to the tidal force from the planet, a satellite on an orbit interior to the synchronous orbit will gradually migrate inward, while the one exterior to the synchronous orbit will migrate outward.
Such orbital evolution will be important when considering the origin and evolution of ring-moon systems.

In the case of rings with sufficiently large surface density and sufficiently inelastic particles, collective gravitational effects between particles become important at a radial location sufficiently far from the central body.
If the particles are close enough to the central object, $R_\mathrm{H} \ll  R$ and mutual gravity between particles can be neglected and they maintain a uniform distribution, even if the particles are densely packed.
At a sufficiently distant location where $R_\mathrm{H} \gg R$, gravitational accretion can occur, as we mentioned above.
In the intermediate radial region, particles in dense rings tend to gather together due to mutual gravity, but are stretched by differential rotation, resulting in the formation of shearing tilted wake structures, with individual wakes forming and dissolving on the timescale of an orbital period \citep{Sal1995,Sal2018}.
This non-axisymmetric structure is called a self-gravity wake (or gravitational wake).
The parameter that describes the importance of collective gravity is called the Toomre $Q$ parameter \citep[see, e.g.,][]{Sch2009}.
In the case of a disk of particles, it is defined by

\begin{equation}
{\displaystyle
Q = \frac{c_r \kappa}{3.36 G \Sigma}.
}
\end{equation}

\noindent
Here, $c_r$, $\Sigma$, $\kappa$ represent the radial velocity dispersion, the surface density of the ring, and the epicyclic frequency ($\kappa = \Omega$ in the case of Keplerian rotation), respectively. 
The ring is gravitationally unstable with respect to axisymmetric perturbations when $Q < 1$, while it is stable if $Q>1$.
When $Q \gtrsim 2-3$, the spatial distribution of particles in the stable ring is almost random.
However, if the surface density is large and/or the distance from the central object is sufficiently large so that $Q \simeq 1-2$, self-gravity wakes are formed in the ring.
In this case, the value of $\tau$ also depends on the direction of the observer, thus, information about fine structures of the ring can be obtained from observation of the optical depth (Section 3, Figure \ref{fig:saturn2}).

Orbits of ring particles can also be disturbed by gravitational perturbations from satellites.
Three natural orbital frequencies can be defined for particles orbiting a central body; 
azimuthal frequency (Keplerian angular velocity in the case of a purely Keplerian motion) related to orbital motion about the central body, radial frequency or epicyclic frequency related to the radial oscillation of the particle, and vertical frequency for the particle's vertical oscillation due to its inclined orbital plane.
If the central body has a flattened shape, the mass distribution within it is not in perfect spherical symmetry, which causes particles orbiting the central body to have three different frequencies.
Effects of perturbation by a satellite becomes prominent at radial locations where the ratio of the frequency of the perturbation to one of the natural frequencies of the particles is related by a ratio of small integers, which is called resonance.
For example, the $2:1$ mean motion resonance occurs when the orbital period of a satellite located radially exterior to the orbit of a ring particle is twice the orbital period of the particle.
The resonant effects of the satellite play an important role in the formation of various types of structure in rings (Section 3).

Collisions and gravitational interactions between particles not only influence velocity dispersion but also lead to spreading of the rings in the radial direction, and the rate of this radial spreading is expressed in terms of viscosity.
In dilute rings with the optical depth $\tau \ll 1$ consisting of non-gravitating particles, the viscosity is on the order of $R^2\Omega \tau$ \citep[see, e.g.,][]{Sch2009}, but in dense rings of self-gravitating particles, the viscosity drastically increases in proportion to the square of the ring surface density \citep{Dai2001}.
This strong dependence of the viscosity on the surface density plays an important role in the structure formation in the rings as well as the global evolution of the ring surface density  \citep[][see Section \ref{chap_ring:sec4}]{Salm2010}.

Jupiter's rings consist of micron-sized dust particles.
Neptune's rings also contain a significant amount of dust, and there are dust rings around Saturn and Uranus.
In these dust rings, different effects are important than those in rings made of macroscopic particles.
For example, a dust particle orbiting a central body absorbs and re-radiates radiation from the Sun, leading to orbital decay toward the central body. This effect is called Poynting-Robertson drag, and is important in dynamical evolution of dust rings. 
In the case of dust rings around giant planets, interactions of charged particles with the magnetic field of the planet are also important.
These influence the orbital evolution and the structure formation of dust rings (Section 3).

\section{Overview of each ring system}\label{chap_ring:sec3}%

\subsection{The rings of Saturn}\label{chap_ring:subsec31}

Galileo Galilei was the first to observe Saturn and its rings by a telescope in 1610. 
Christiaan Huygens suggested that the rings are thin and disk-shaped, and James Clerk Maxwell proved that they are composed of countless small particles orbiting Saturn on individual orbits. 
Observations of Saturn, its moons and rings continued using ground-based telescopes, but the arrival of spacecraft made a great leap forward in our understanding of the system.
The first spacecraft to visit Saturn was Pioneer 11 (1979).
The two spacecraft, Voyager 1 (1980) and Voyager 2 (1981), equipped with higher-resolution cameras than Pioneer 11, have made many new discoveries about Saturn and its moons and rings during their flybys. 
Then, Cassini spacecraft for the first time entered into orbit around Saturn in July 2004.
After its four-year prime mission, the mission was extended twice, and the spacecraft continued observations until September 2017.
As the spacecraft was running out of fuel, the mission was intentionally ended by crashing into Saturn's atmosphere to avoid contamination of potentially habitable moons.

\begin{figure}[bth]
\centering
\includegraphics[bb=0.000000 0.000000 2671.670000 969.880000, width=0.95\textwidth]{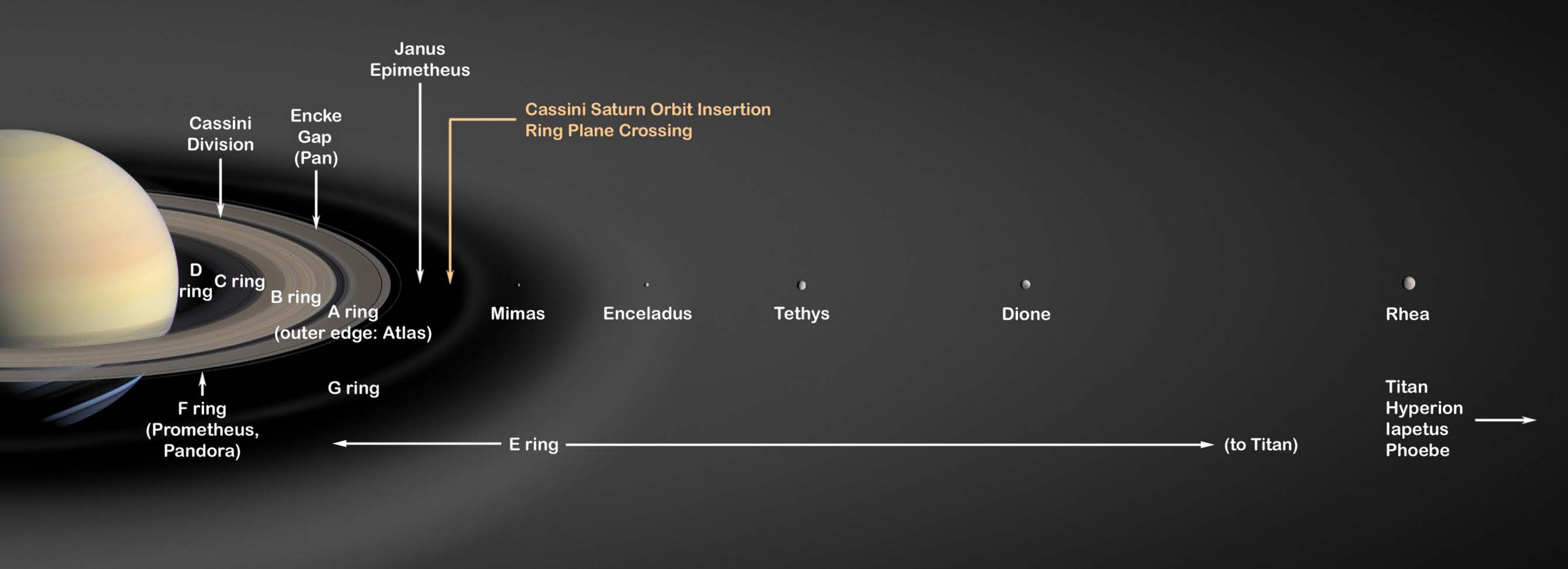}
\caption{An artist's concept of Saturn's rings and major icy moons. 
The seven rings, D, C, B, A, F, G and E from the planet outward, are named in the order of their discovery. The diffuse E ring extends from Mimas' orbit to the orbit of Titan, which has an orbital radius about 20 Saturn radii and is not shown here. 
The radial location of the ring plane crossing of the Cassini spacecraft at the time of its Saturn orbit insertion is also indicated. (NASA/JPL; Part of the inserted labels has been modified with assistance from Takaaki Takeda.)}
\label{fig:saturn1}
\end{figure}

Saturn's rings are divided into several parts and labeled as A through G in the order in which they were discovered (Figure \ref{fig:saturn1}). 
The brightest part that can be observed from Earth even by a small telescope are the A and B rings, and the dark area between these rings is called the Cassini Division, after its discoverer Giovanni Cassini. 
Inside the B ring is the faint C ring, and these three are called the main rings. In fact, the Cassini Division is not a complete gap, but a dilute ring similar to the C ring. 
Inside the C ring is the even fainter D ring. 
On the other hand, outside the main rings, the E ring extends from the orbit of the moon Mimas to the orbit of Titan, which have an orbital radius about 3 and 20 Saturn radii, respectively. 
Also, there is the narrow F ring just outside the main rings, and the faint G ring between the F ring and the inner edge of the E ring.

Saturn's main rings have a diameter of about $2.7 \times 10^5$ km, which is about 70\% of the distance from the Earth to the Moon, but are very thin, typically about 10 m thick. This thickness to width ratio on the order of $10^{-7}$ means that Saturn's rings are as thin as a sheet of 
tissue paper 
spread across a football field, and are virtually invisible when viewed edge-on from Earth. 
During the last phase of its mission, Cassini passed between the inner edge of the D ring and Saturn's cloud top, enabling the measurement of the ring mass.
The results showed that the total mass of the rings is $1.5 \times 10^{19}$ kg, about 0.4 times the mass of the moon Mimas \citep{Ies2019}.
Saturn's rings are mainly composed of water ice. The particles that make up the main rings are typically a few centimeters to 10 m in size, while the E ring is a ring of micron-sized dust particles, and the G ring is also mainly composed of dust.

\begin{figure}
\begin{center}
\begin{tabular}{ll}
\includegraphics[bb=0.000000 0.000000 435.000000 230.000000, height=0.25\textwidth,clip]{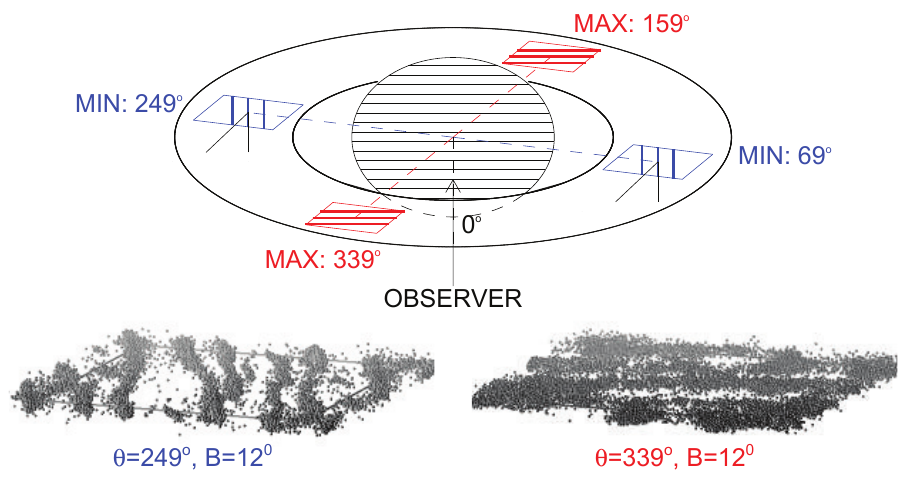}
&
\hspace{3mm}
\includegraphics[bb=0.581131 0.000000 301.898000 266.000000, height=0.40\textwidth,clip]{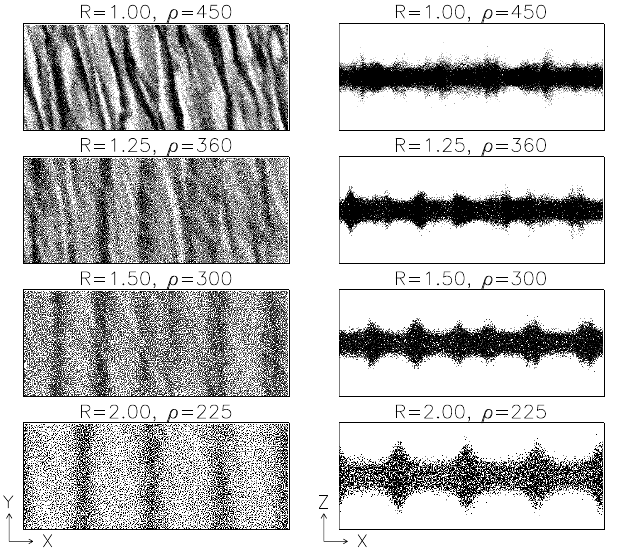}
\end{tabular}

\caption{
Left: Schematic explanation for the relation between self-gravity wake structures and azimuthal brightness asymmetry in Saturn's rings (from \citet{Sal2004}; see also \citet{Sch2009}). At low viewing angle 
$B$
from the ring plane, the wakes trailing by about $21^{\circ}$ with respect to the local tangential direction are seen roughly along their long axis at ring longitudes of  $249^{\circ}$ and $69^{\circ}$, and perpendicular to their long axis at  longitudes of  $339^{\circ}$ and $159^{\circ}$. In the former case rarefied regions between wakes are visible, reducing the reflecting area and corresponding to minimum brightness. In the latter case, the rarefied regions are hidden by the wakes, maximizing the reflecting area and corresponding to maximum brightness.
Right: Snapshots from self-gravitating simulations of a local rectangular region of 583 m × 233 m in rings at 100,000 km from Saturn (from \citet{Sal2001}; see also \citet{Sch2009}). 
The four different examples correspond to different combinations of particles' internal density $\rho$ (in kg m${}^{-3}$) and radius $R$ (in meters): To maintain fixed surface density and optical depth of the rings, the product $R \times \rho$ is kept fixed. In the left column the system is shown from above (the planet is to the left, and the direction of the orbital motion is up), while in the right column the system is shown from the side.
Transition from wake-dominated structures (upper row) to overstability-dominated structures (bottom row) can be seen.
}
\label{fig:saturn2}
\end{center}
\end{figure}

Particles in the rings repeatedly collide and bounce off without merging. 
However, at sufficiently dense part of the rings located far enough away from the planet, the particles tend to gather together due to their mutual gravity. 
The particles are then stretched in the direction of the orbital motion due to the shear motion arising from the difference in the orbital angular velocity, and elongated wake structures with a spacing of about 100m are formed (Figure \ref{fig:saturn2}, left panel and right-top panel). 
These structures are called gravitational wakes or self-gravity wakes.
Such structures in planetary rings were first demonstrated in a computer simulation that includes collision and gravitational interaction between particles \citep[see][for details]{Sal1995,Sal2018}.
Self-gravity wakes can explain the observed azimuthal brightness variation of Saturn's rings, because the reflecting surface area is the largest and the part of the rings appears brightest when the wakes are seen edge-on, while the area is the smallest and appears less bright when seen along the long axis of the wakes (Figure \ref{fig:saturn2}, left panel).
Gravitational wake structures are too small to be directly seen even with the resolution of the images taken by Cassini. 
However, the Hubble Space Telescope and multiple instruments onboard Cassini indirectly confirmed these structures in the A and B rings by occultation observations, where the star light partly blocked by the rings is observed from various angles.

On the other hand, several instruments onboard Cassini found in the A and B rings quasi-periodic radial density variations with characteristic wavelengths of a few hundred meters. The most promising explanation for such axisymmetric structures is viscous overstability, which is an oscillatory instability that occurs through a coupling of the viscous stress to the background orbital shear \citep[][Figure \ref{fig:saturn2}, right panels]{Sch2001,Mon2023}.

\begin{figure}
\begin{center}
\begin{tabular}{ll}
\begin{tabular}{l}
\includegraphics[bb=0.000000 0.000000 707.000000 730.000000, width=0.24\textwidth,clip]{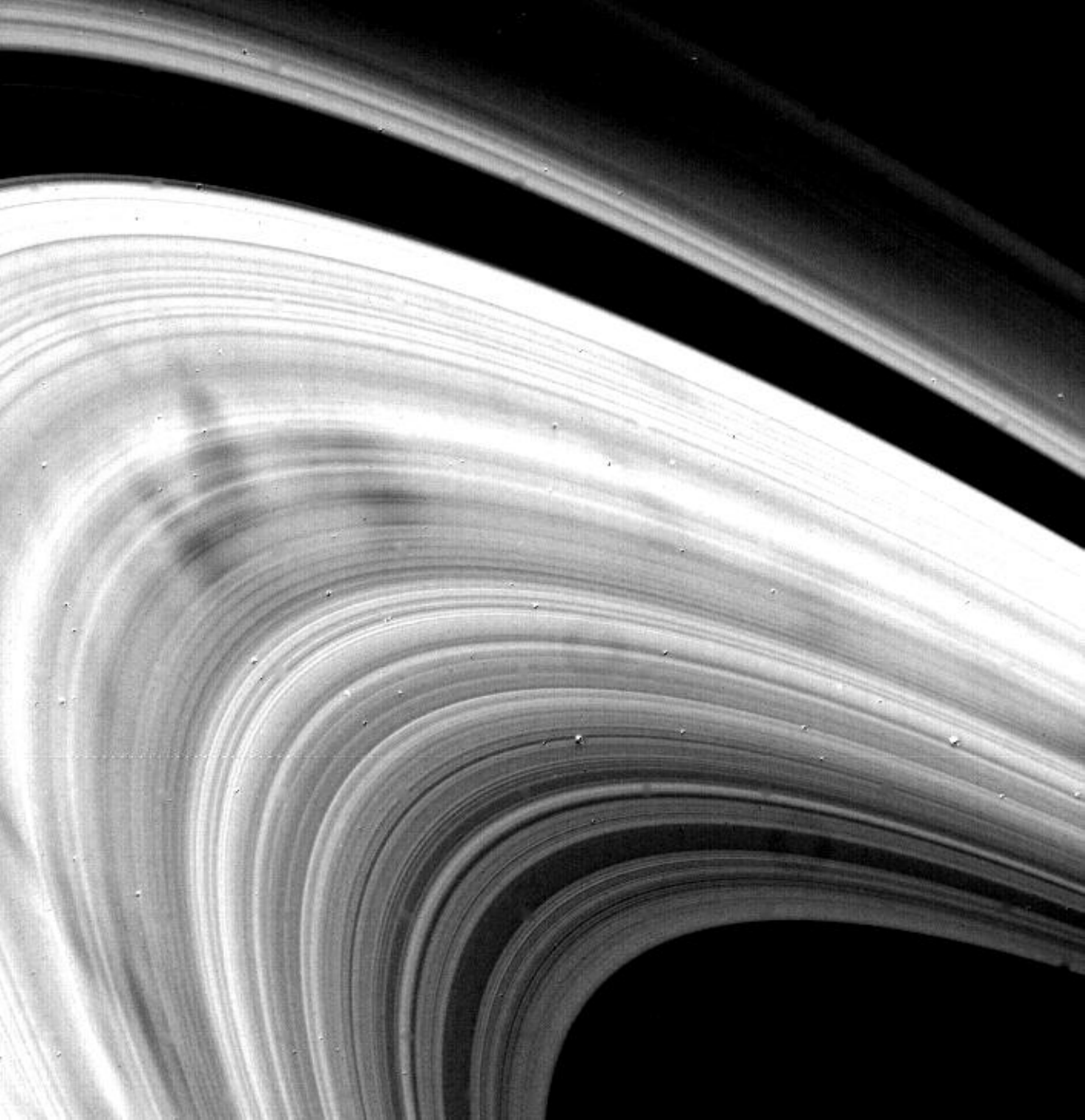}\\
\includegraphics[bb=0.000000 0.000000 1008.000000 1008.000000, width=0.24\textwidth,clip]{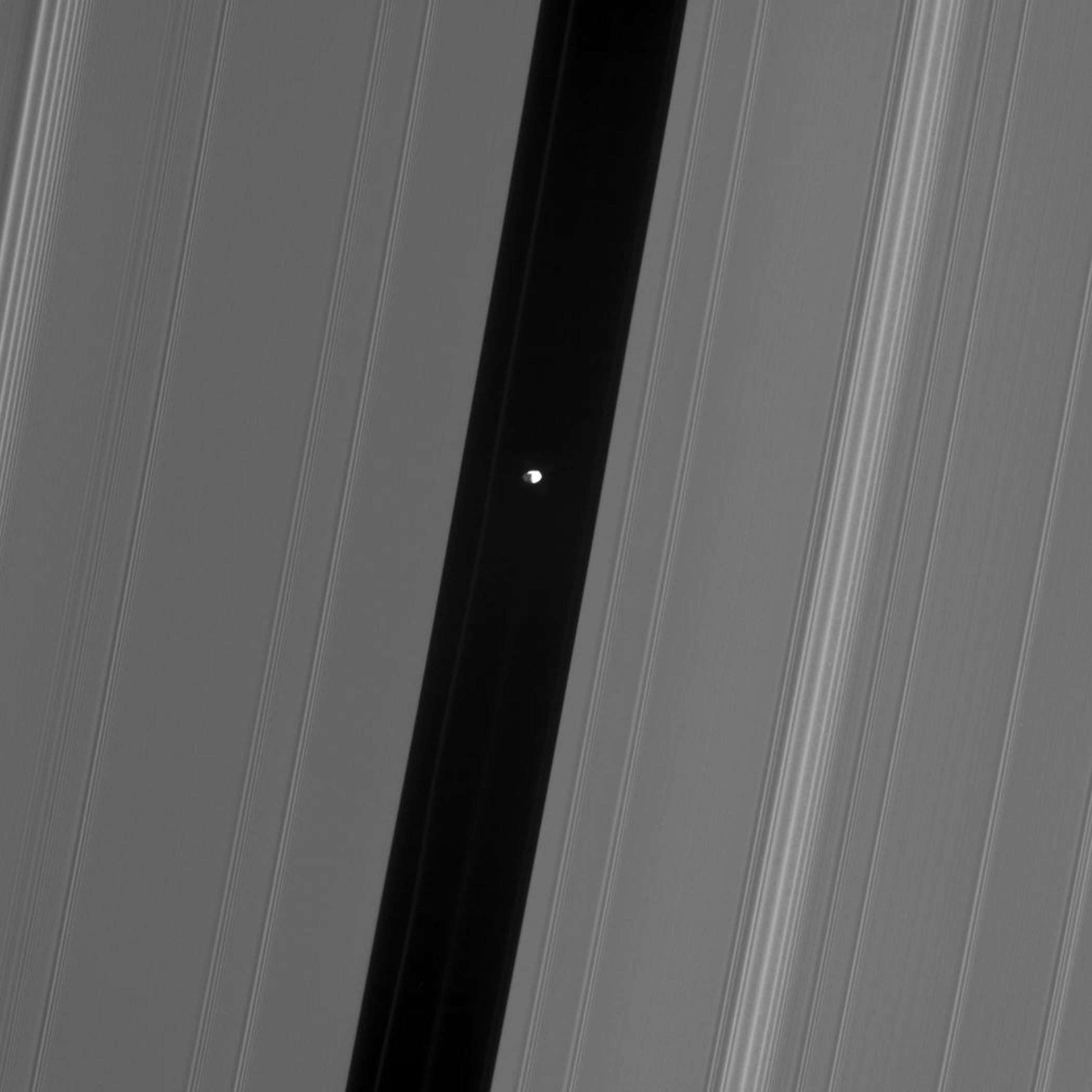}
\end{tabular}&
\hspace{-8mm}
\begin{tabular}{l}
\includegraphics[bb=0.000000 0.000000 1024.000000 1024.000000, width=0.226\textwidth,clip]{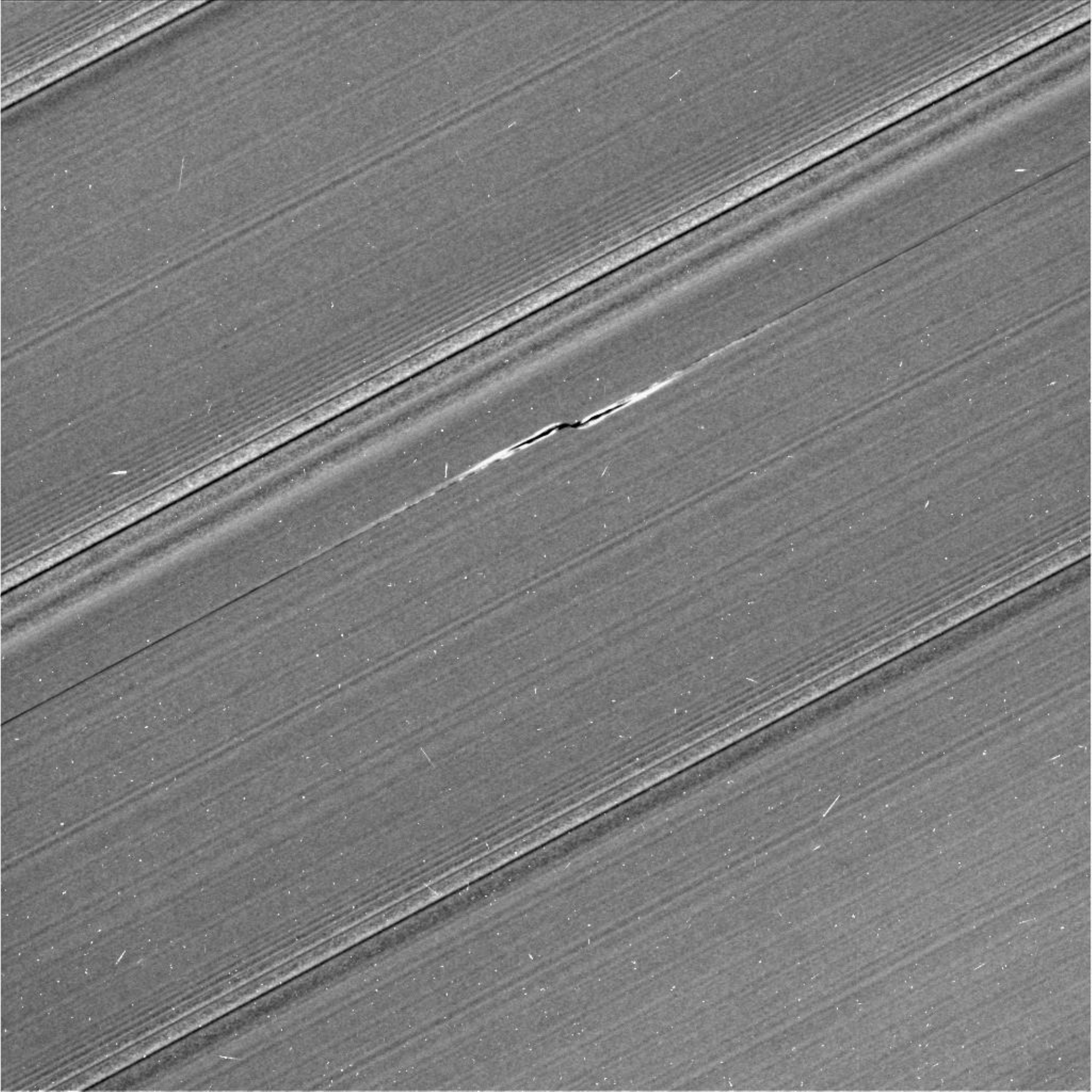}\\
\includegraphics[bb=0.000000 0.000000 1056.000000 1224.000000, width=0.226\textwidth,clip]{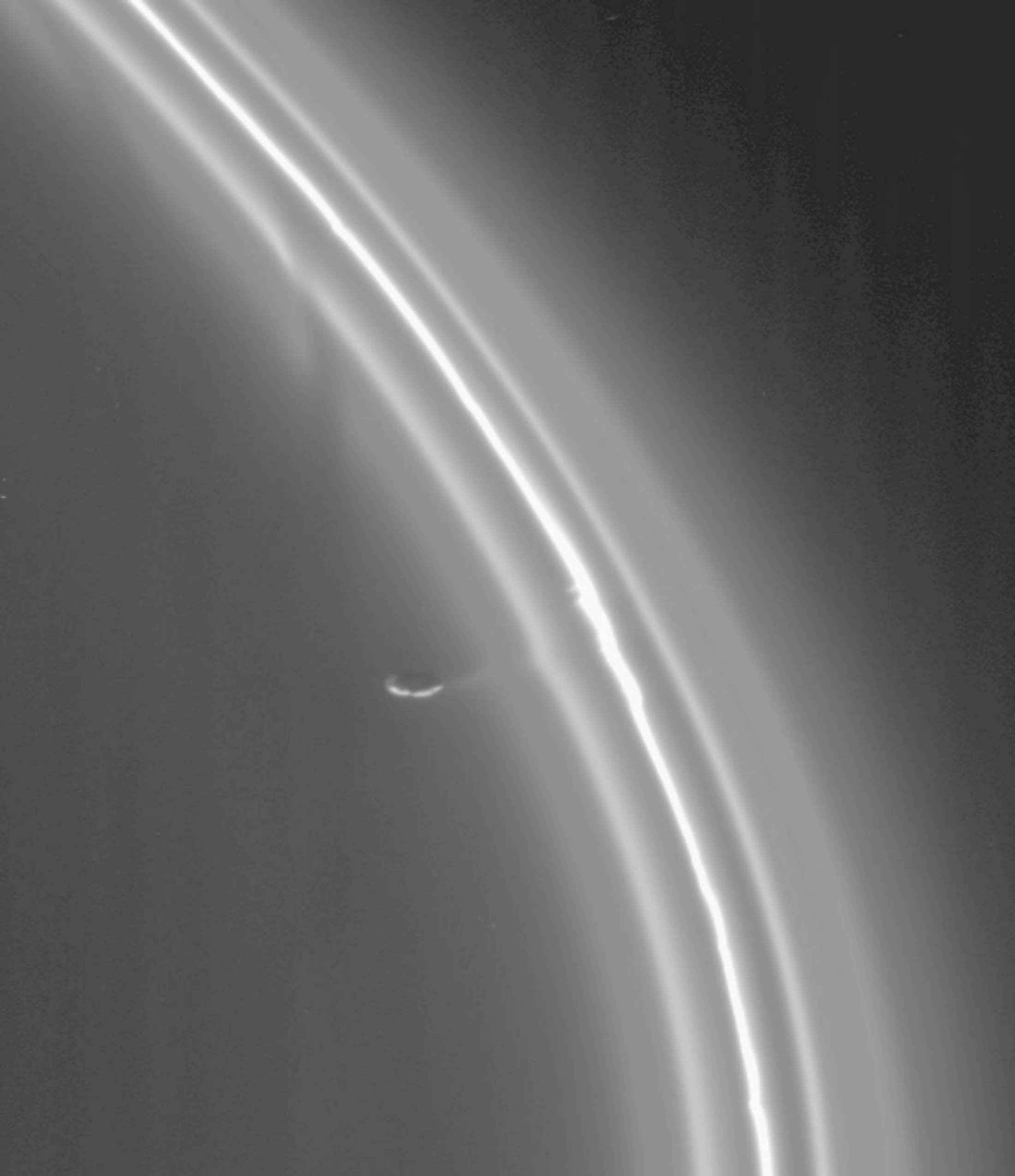}
\end{tabular}
\end{tabular}
\caption{
Left-top: Spoke feature in the B-ring observed by Voyager 2. 
Left-bottom: Pan in the Encke gap within the A ring.
Right-top: Image of the propeller feature in the A ring known informally as Bleriot.
Right-bottom: F ring and Prometheus. 
These three images were obtained by Cassini
(NASA/JPL/Space Science Institute).
}
\label{fig:saturn3}
\end{center}
\end{figure}

Furthermore, images from the Voyager spacecraft showed radial features in the B ring, which are called spokes (Figure \ref{fig:saturn3}, left-top panel). The spokes contain a large amount of micron-sized dust particles, and first appear as radial patterns and rotate with Saturn's magnetic field, then they
become stretched in the direction of orbital motion by the Keplerian differential rotation of the particles \citep{Hor2009}.
The electromagnetic effects acting on dust particles are likely to be responsible for the formation of the spokes, but the mechanism for the formation is still not well understood.

\begin{figure}
\begin{center}
\begin{tabular}{ll}
\begin{tabular}{l}
\includegraphics[bb=0.000000 0.000000 535.000000 357.000000, width=0.24\textwidth,clip]{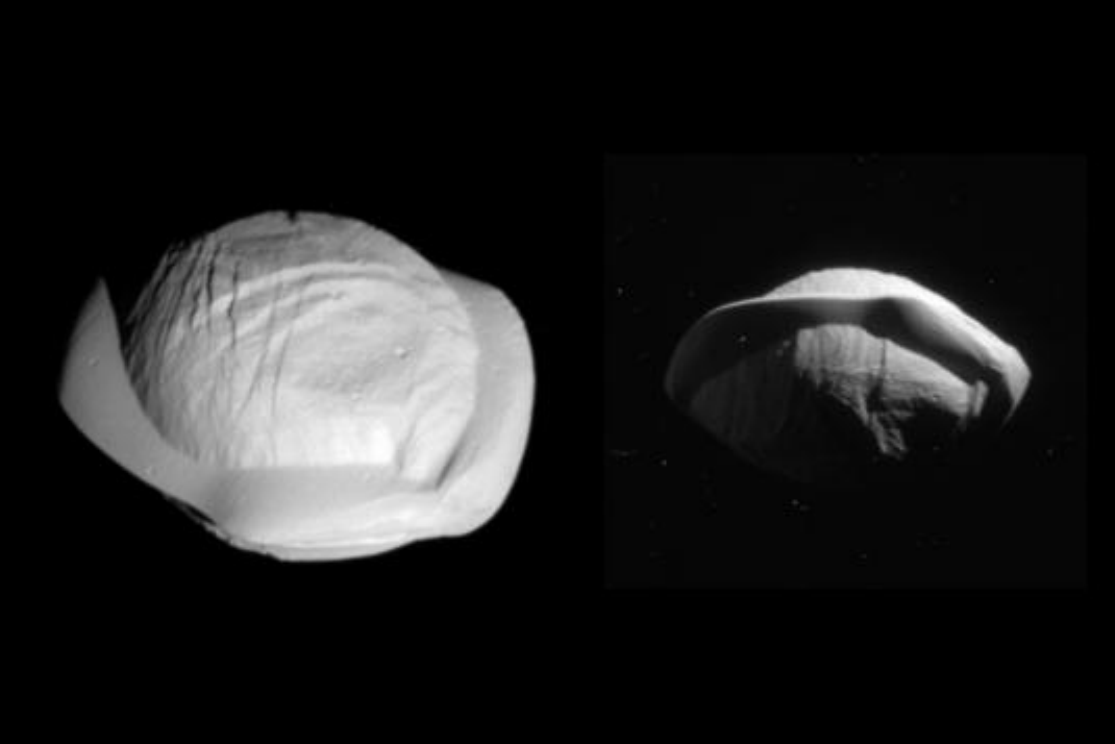}\\
\includegraphics[bb=0.000000 0.000000 1024.000000 1024.000000, width=0.24\textwidth,clip]{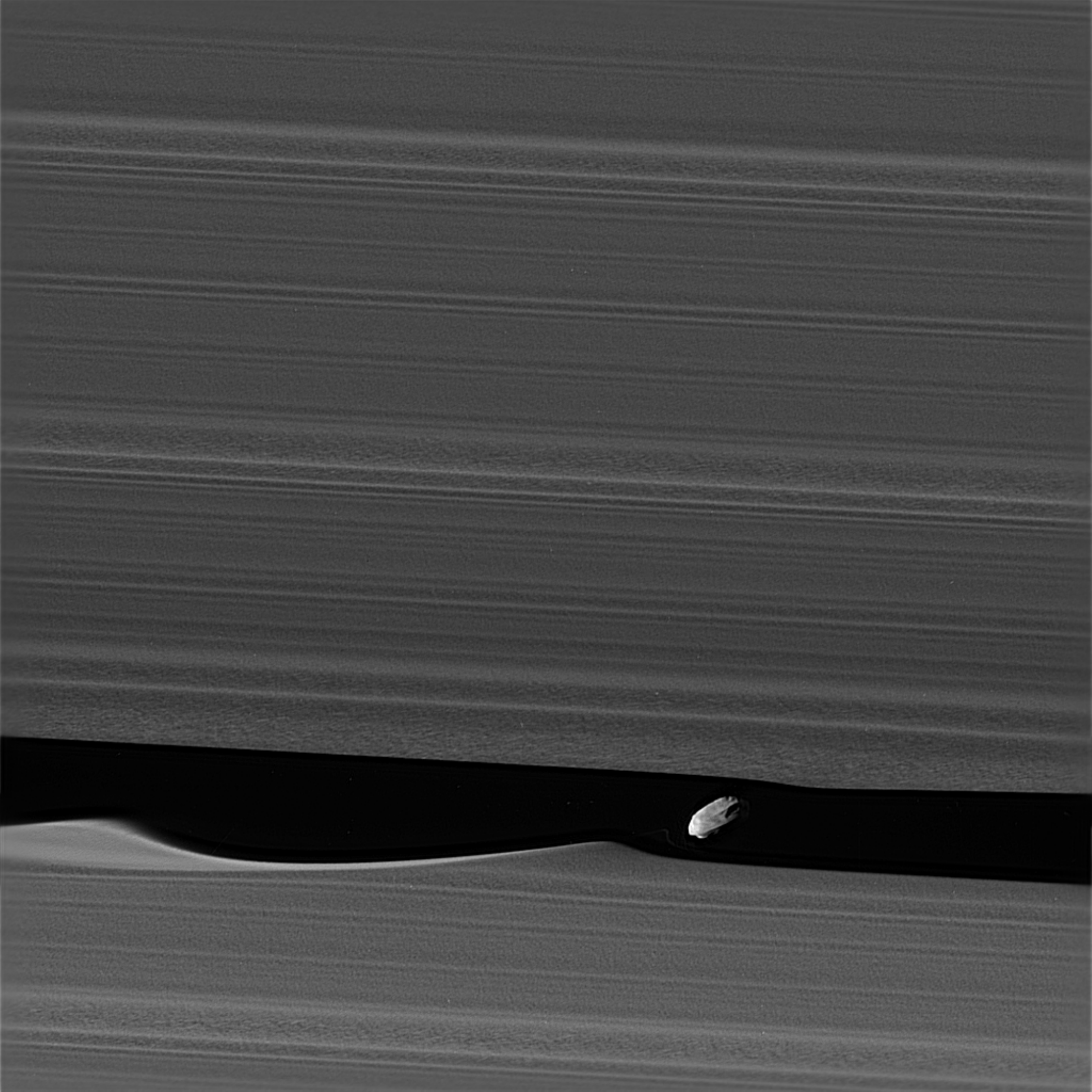}
\end{tabular}&
\hspace{-8mm}
\begin{tabular}{l}
\includegraphics[bb=0.000000 0.000000 768.000000 516.000000, width=0.264\textwidth,clip]{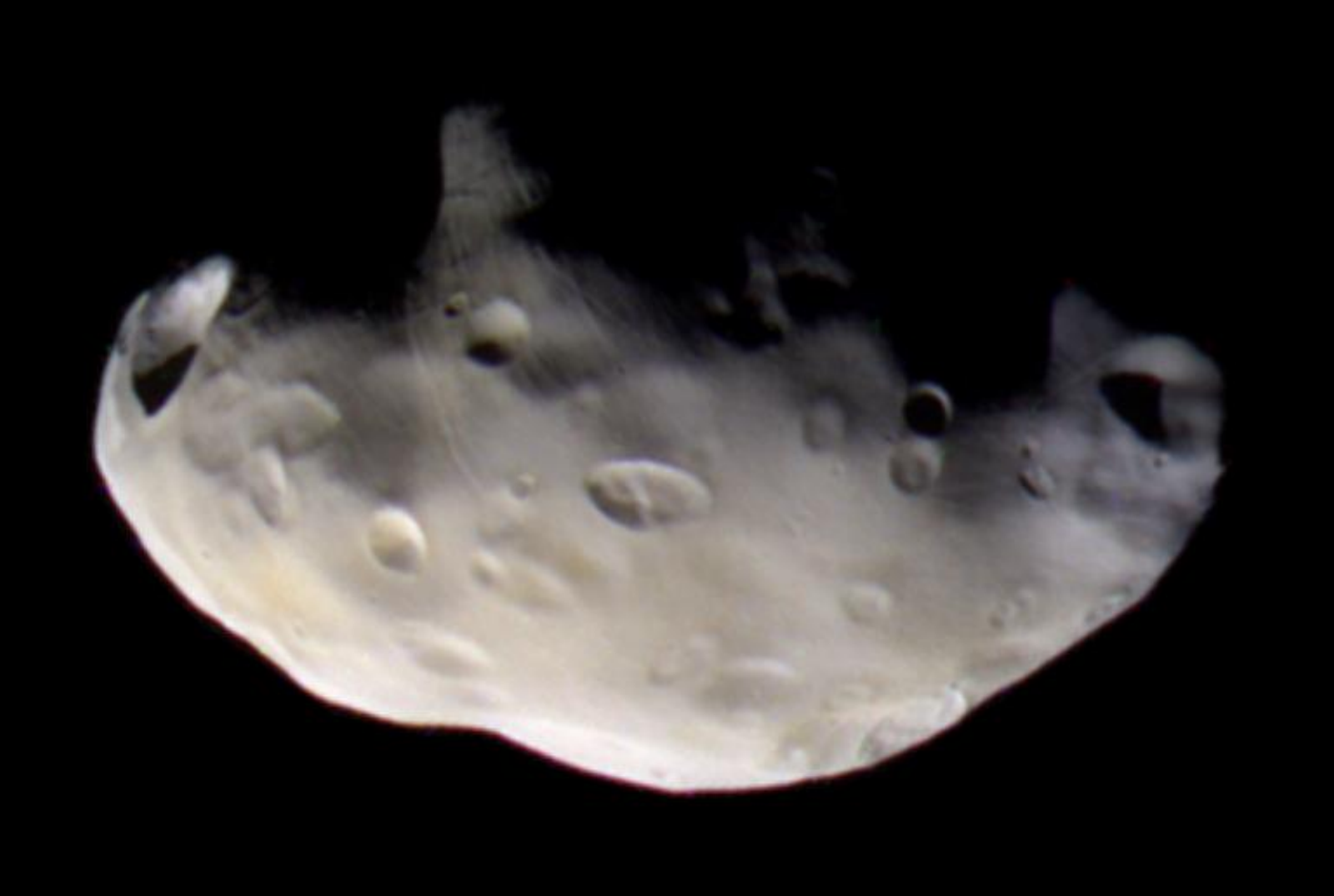}\\
\includegraphics[bb=0.000000 0.000000 1019.000000 863.000000, width=0.264\textwidth,clip]{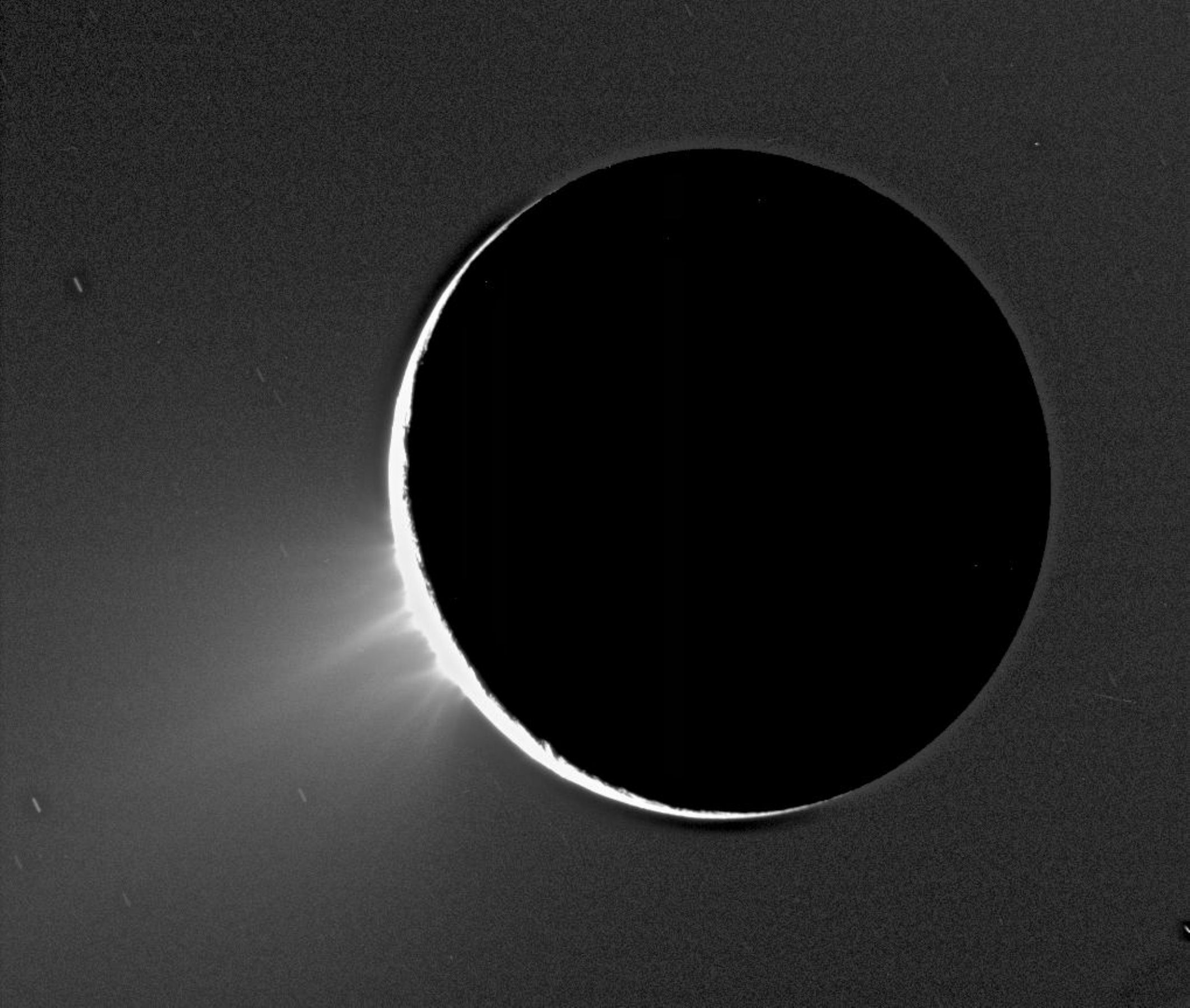}
\end{tabular}
\end{tabular}
\caption{
Left-top: Images of the northern and southern hemispheres of Pan, at left and right, respectively.
Left-bottom: Daphnis within the Keeler Gap in the A ring.
Right-top: Pandora, a moon orbiting just outside the F ring.
Right-bottom: Enceladus with plumes erupting from its surface, which is making up the E ring. 
All these images were taken by Cassini (NASA/JPL/Space Science Institute).
}
\label{fig:saturn4}
\end{center}
\end{figure}

A sufficiently massive moon embedded in rings can create a complete annular gap by gravitational scattering of nearby particles.
The Encke Gap in the A ring is created by a small moon Pan (Figure \ref{fig:saturn3}, left-bottom panel).
Unresolved images of Pan were found in 1990 in the Voyager 2 archive after careful search. 
Cassini took high-resolution images of the Encke Gap and Pan, and revealed that Pan has a peculiar shape like a flying saucer  (Figure \ref{fig:saturn4}, left-top panel). 
Further out from the Encke gap and near the outer edge of the A ring, there is the narrower Keeler Gap.
Cassini revealed detailed structures of the edge of the Keeler gap, and discovered the small moon Daphnis, which creates this gap  (Figure \ref{fig:saturn4}, left-bottom panel). 
The passage of Daphnis induces wavy structures at the gap edges, leading wakes at the inner edge where the particles move faster than the moon, and trailing wakes at the outer edge where the particles move slower \citep{Tor2024}.
From the images taken around the time of Saturn's equinox, the vertical scale of the structure is estimated from its shadow cast on the ring plane to be more than 1 km.

Even if not massive enough to create a complete gap, a moon significantly more massive compared to the surrounding ring particles can create a local structure by gravitational scattering of surrounding particles.
As in the case of the wavy structures at the gap edges, leading and trailing wakes are created interior and exterior to the moon's orbit, respectively, forming the so-called propeller structure.  
Formation of such structure around an embedded moonlet was theoretically predicted, and was confirmed in high-resolution images of the A ring taken by Cassini, although the embedded moonlet itself is too small to be directly imaged  \citep[][Figure \ref{fig:saturn3}, right-top panel]{Tis2006}.
From the size of the propeller structure, the typical size of the embedded moonlet that creates the structure is estimated to be on the order of 100 m. 
Sufficiently large propeller features have been confirmed in multiple images, which allows tracking of their orbital evolution, likely caused by interaction with surrounding particles. 
Such large propeller features are informally named after aviators, such as ''Bleriot'' after Louis Bl\'{e}riot, who was the first person to fly across the English Channel.

In the outer regions of the main rings where the radial distance from the planet approaches the Roche limit $a_\mathrm{Roche}$, the Hill radius of ring particles becomes comparable to their physical radius.
When the Hill radius exceeds the physical radius of the particles, colliding particles with sufficient energy dissipation become bound due to mutual gravity, leading to the formation of gravitational aggregates and accretion of satellites.
Near the outer edge of the A ring where the Hill radius of a body barely exceeds its physical radius, an aggregate formed by gravitational accretion of particles is expected to have a lemon-shape, reflecting the shape of the Hill sphere.
Images of small moons near the outer edge of the A ring taken by Cassini showed that these moons in fact have elongated shapes like the Hill sphere, implying that they were formed by gravitational accretion of small particles onto a core or accretion and reshaping after a collision between small moons \citep[Figure \ref{fig:saturn4}, right-top panel;][]{Por2007,Lel2018}.

Moons near the outer edge of the A ring and further away exert gravitational influence on the rings and form various structures. 
For example, the inner edge of the Cassini Division is governed by a strong 2:1 mean motion resonance with the moon Mimas.
Also, the outer edge of the A ring is maintained by the overlapping effects of resonances from several nearby moons.
At a Lindblad resonance, where the gravitational perturbation from the satellite is in resonance with the radial frequency of a ring particle, eccentricities of the particles at the location of the resonance are enhanced, resulting in the formation of various structures.
Lindblad resonances are responsible for producing spiral density waves in the rings.
Since the change in the wavelength as the wave propagates through the ring depends on the surface density of the rings, observation of the density waves provides us information about the mass of the ring. 
In addition to gravitational perturbation form satellites, detailed analysis of data obtained by Cassini revealed waves in the rings induced by vibrations within Saturn, which can provide constraints on the internal structure and rotation rate of the planet \citep{Hed2013,Man2020}.

The F, G, and E rings, which are located exterior to the main rings, are each associated with moon(s) in unique ways. 
The F ring was discovered by Pioneer 11, and Voyager found the ring to accompany two small moons, Pandora and Prometheus, and have a strange twisted shape.
Images taken by Cassini revealed that the ring has a core ring consisting of relatively large particles, and a surrounding more dusty spiral strand (Figure \ref{fig:saturn3}, right-bottom panel). 
The F ring may have formed when the two nearby moons Prometheus and Pandora collided in the past and resulted in partial disruption \citep{Hyo2015}. 
Further out from the F ring is the faint G ring, which has a relatively bright arc. 
Cassini discovered a small moon Aegaeon near the center of the arc. 
The G ring is thought to be made of dust particles generated by meteoroids colliding with this small moon and the surrounding particles in the arc.
The E ring is a dusty ring that extends over a wide radial region further out (Figure \ref{fig:saturn1}). One of the most striking discoveries by Cassini was jets of ice and gas erupting from the south-polar region of Enceladus \citep[][Figure \ref{fig:saturn4}, right-bottom panel]{Por2006}. 
The ice particles erupting from inside Enceladus are forming the E ring.

In 2009, using NASA's infrared Spitzer Space Telescope, a faint ring was discovered just inside the orbit of the moon Phoebe, which orbits Saturn at an average distance of 215 Saturn radii \citep{Ver2009}. 
The ring is called the Phoebe ring, and extends from Phoebe's orbit inward to the orbit of Iapetus, whose average orbital radius is 59 Saturn radii.
The ring is thought to have been formed by particles ejected at meteoroid impacts on Phoebe.

\subsection{The rings of Uranus}\label{chap_ring:subsec32}

\begin{figure}[htb]
\centering
\begin{minipage}[b]{0.48\columnwidth}
\centering
\includegraphics[bb=0.000000 0.000000 1046.120000 1237.350000, height=1.2\textwidth]{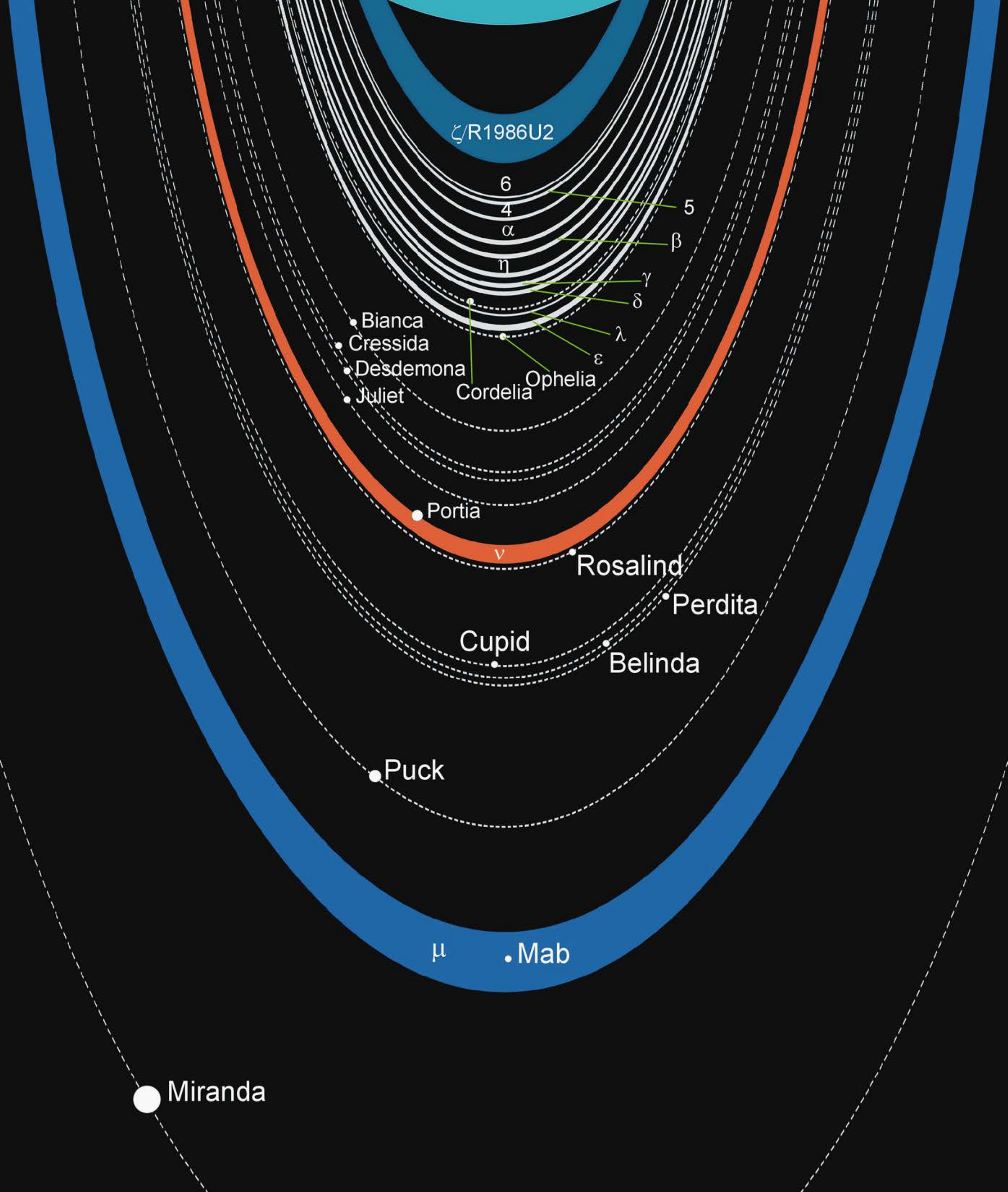}
\caption{Schematic illustration of the Uranian ring-moon system. Solid lines denote rings, and the dashed lines denote orbits of moons. The colored bands show approximate radial extent of the dusty rings (Wikimedia).}
\label{fig:uranus1}
\end{minipage}
\hspace{0.02\columnwidth}
\begin{minipage}[b]{0.48\columnwidth}
\centering
\includegraphics[bb=0.000000 0.000000 800.900000 1516.810000, height=1.05\textwidth]{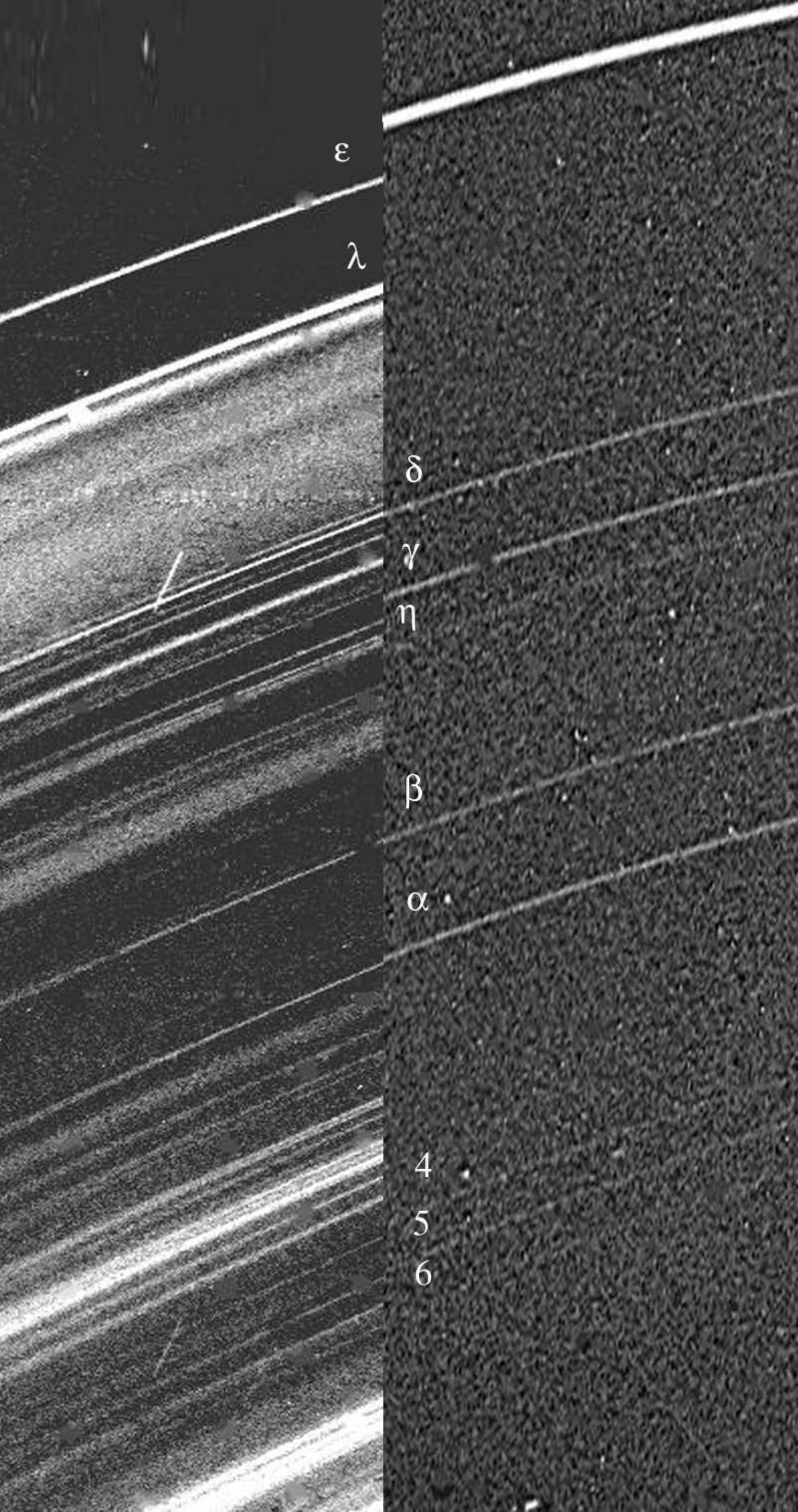}
\caption{Merging two images of Uranian rings obtained by Voyager 2 spacecraft. 
The left panel is the forward-scattering image, showing dust particles in the ring system, while the right panel is the back-scattering image, showing the distribution of larger particles. The $\alpha$ and $\beta$ rings are matched between the two images. The difference in the position of the $\varepsilon$ ring in the two images is caused by its non-zero eccentricity (NASA and Wikimedia Foundation).}
\label{fig:uranus2}
\end{minipage}
\end{figure}

\begin{figure}[htb]
\centering
\includegraphics[bb=0.000000 0.000000 1268.000000 1268.000000, height=0.3\textwidth]{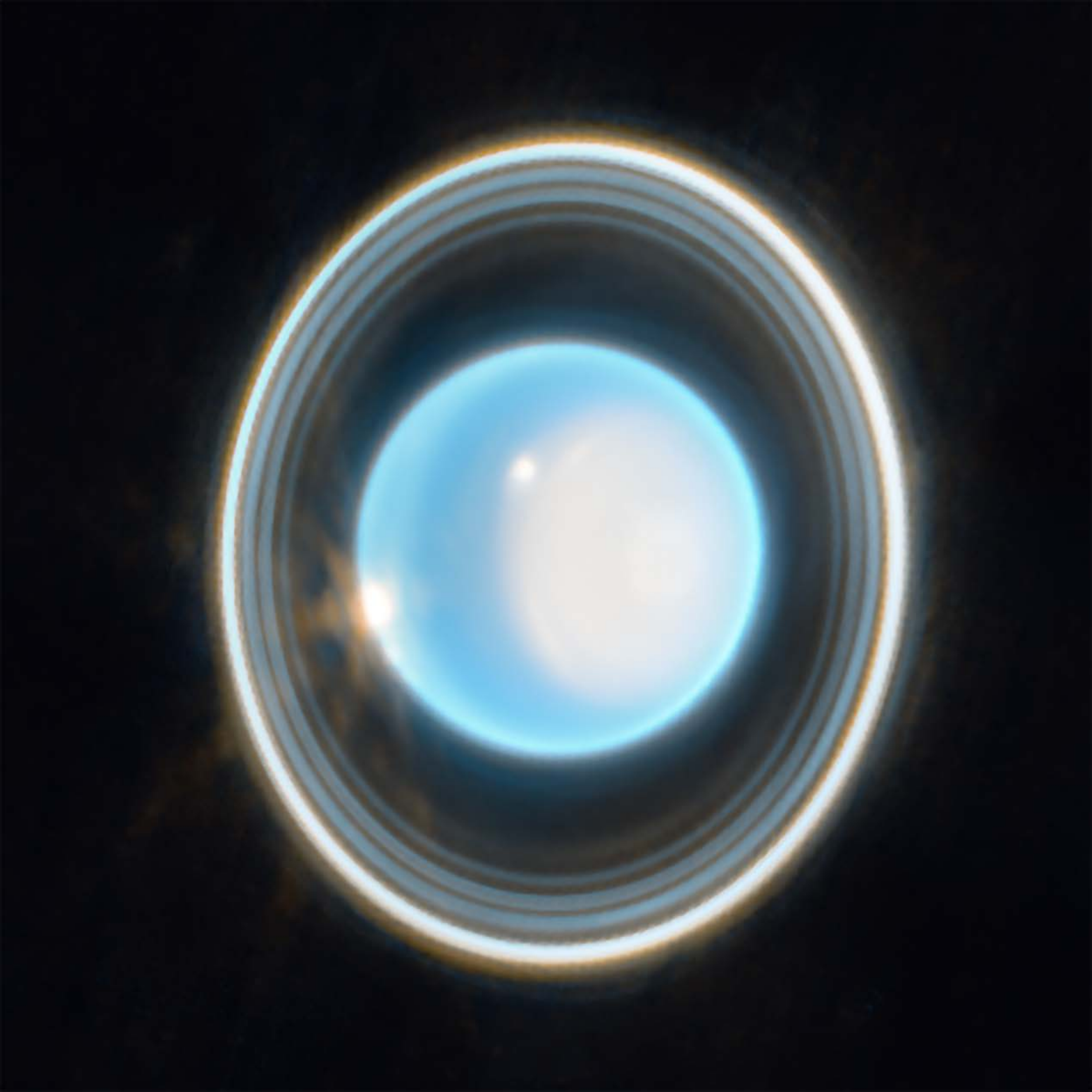}
\caption{Image of the ring system of Uranus taken by the James Webb Space Telescope's Near-Infrared Camera in 2023. 
Uranus rotates on its side, with its axis of rotation being approximately parallel to its orbital plane.
In this image, the northern hemisphere is visible and is facing the Sun.
The brightest ring shown in this image is the $\varepsilon$ ring (NASA/ESA/STScl/James Webb Telescope).}
\label{fig:uranus3}
\end{figure}

The rings of Uranus were discovered in 1977 when astronomers observed Uranus passing in front of a distant star, a phenomenon called stellar occultation  \citep{Ell1977}.
The observation was performed at the Kuiper Airborne Observatory, an aircraft equipped with a telescope, and the observation was expected to provide information about the planet's size and atmosphere.
But they found that the star light was unexpectedly blocked several times before and after it was eclipsed by the planet, leading to the serendipitous discovery of the rings.
This made Uranus the second ringed planet after Saturn.
From the initial analysis of the data obtained by this observation, five rings, $\alpha$, $\beta$, $\gamma$, $\delta$, and $\varepsilon$ rings, were confirmed.
Further detailed analysis of the data obtained by this observation and other data confirmed three additional rings (rings 6, 5, and 4) inside the $\alpha$ ring, and a new ring ($\eta$ ring) between the $\beta$ and $\gamma$ rings, making a total of nine rings (Figure \ref{fig:uranus1}).
Later, in 1986, Voyager 2 directly imaged Uranus' rings (Figure \ref{fig:uranus2}) and discovered two new rings ($\lambda$ and $\zeta$ rings).
Additionally, from 2003 to 2005, observations with the Hubble Space Telescope discovered two new broad, very faint rings $\mu$ and $\nu$ farther away from the planet (see \citet{Nic2018} for a review).
More recently, NASA's James Webb Space Telescope has taken an image of Uranus, including its rings (Figure \ref{fig:uranus3}).

There are two types of rings on Uranus: dense narrow rings and dusty rings with low optical depth (Figure \ref{fig:uranus1}).
There are a total of 10 narrow rings: 6, 5, 4, $\alpha$, $\beta$, $\eta$, $\gamma$, $\delta$, $\lambda$, and $\varepsilon$ from the inside, and most of them are dense rings with non-zero eccentricity and inclination. 
The $\varepsilon$ ring, which is the brightest and densest of these rings, has a notable elliptical shape.
The optical depth of the $\varepsilon$ ring ranges from 0.5 to 2.5, with a large value near the periapsis, where the ring becomes narrower.
The radial positions of the edges of the narrow rings such as $\varepsilon$, $\delta$, and $\gamma$ often match the radial locations of mean motion resonances with nearby satellites such as Cordelia and Ophelia, and these satellites seem to play a role in maintaining the shape of the rings.

On the other hand, there are three main dust rings: $\zeta$, $\nu$, and $\mu$ from the inside (Figure \ref{fig:uranus1}).
Also, the high-resolution image taken by Voyager 2 suggests that there are many dust particles surrounding the narrow rings (Figure \ref{fig:uranus2}).
The narrow $\lambda$ ring is the brightest ring in the forward-scattering image, indicating that this ring is composed primarily of dust particles.
The presence of various radial strutures in the dust rings between the narrow rings may indicate the existence of undiscovered small moons.
The dusty $\nu$ ring, which is located among the orbits of small moons, may have been generated by recent collisions between the moons. 
There is also a satellite, Mab, located in the center of the $\mu$ ring, and this satellite may be involved in the origin of this dust ring.

The albedo of Uranus' rings is very low, thus the particles in the rings are likely to be covered with dark material, but their composition is not known. 
It may reflect some primitive carbonaceous chondritic material, or some kind of organic matter darkened due to radiation.
Many small satellites orbit near the rings of Uranus (Figure \ref{fig:uranus1}).
This suggests that these satellites as well as undiscovered small moons are likely to be involved in the origin of the rings through some mechanisms such as generation of ejecta at meteoroid impacts onto these satellites; disruption, partial disruption, or re-accretion of these satellites and their fragments.
The formation of the whole ring-satellite system may be related to a giant impact that tilted the spin axis of the planet.

\subsection{The rings of Jupiter}\label{chap_ring:subsec33}

\begin{figure}[t]
\centering
\includegraphics[bb=0.000000 0.000000 405.260000 283.420000, width=0.6\textwidth]{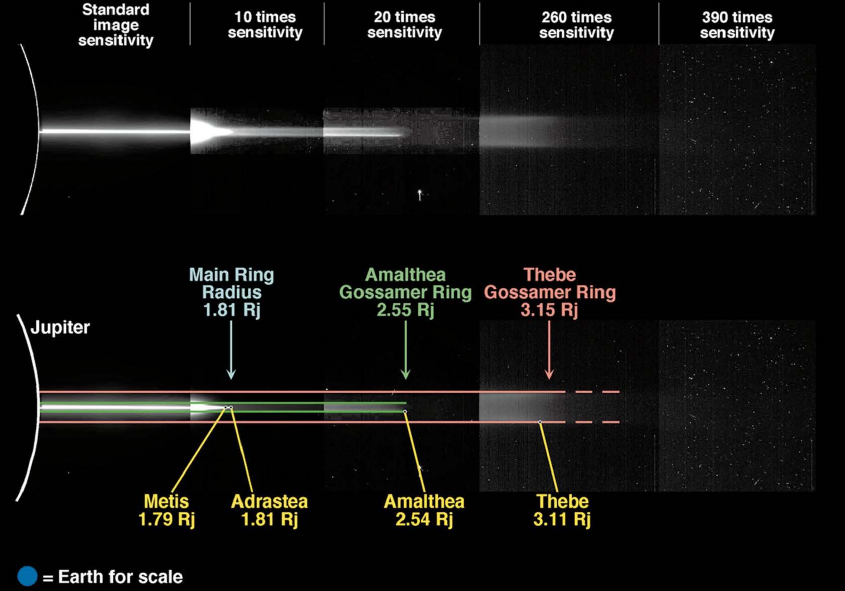}
\caption{Mosaic of the five nearly edge-on images of Jupiter's rings taken by the Galileo spacecraft such that the vertical extent of the rings is showcased. The mosaic is shown twice, on the same scale; the top panel displays only the data, while the bottom panel also gives the location of small moons and presents a match between the image and a simple geometrical model of the gossamer rings ($R_\mathrm{j}$ is Jupiter's radius). Image sensitivity increases from left to right, in order to make increasingly faint structure visible (NASA/\cite{Ock1999}).}
\label{fig:jupiter1}
\end{figure}

\begin{figure}[h]
\centering
\begin{minipage}[b]{0.48\columnwidth}
\centering
\includegraphics[bb=0.000000 0.000000 517.980000 321.810000, width=0.9\textwidth]{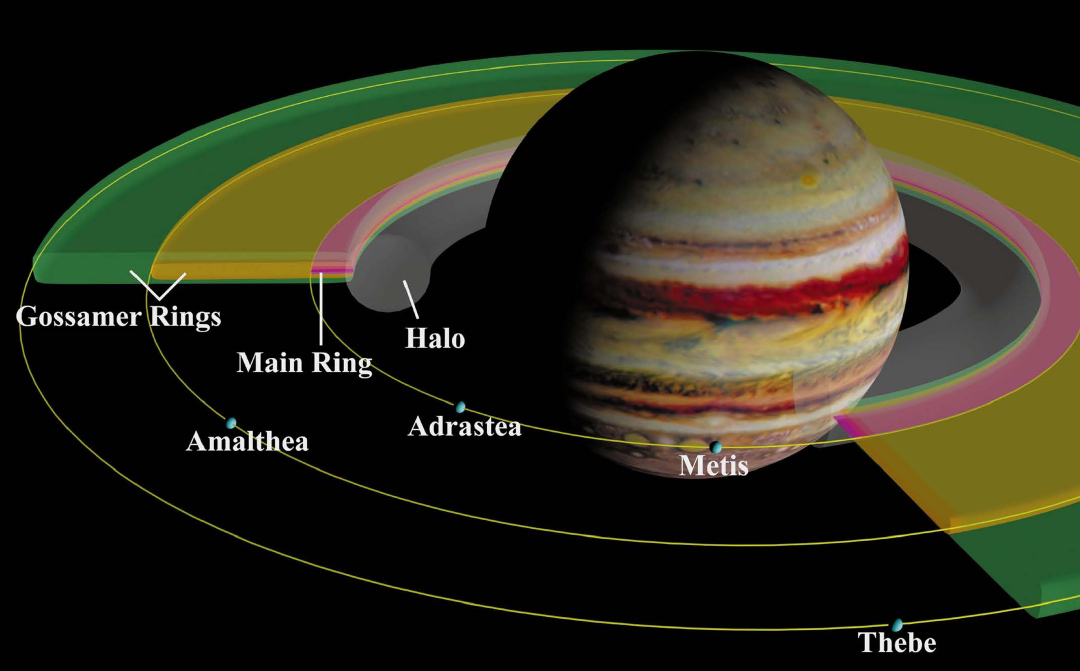}
\caption{Schematic illustration showing the relationship between Jupiter, its rings, and the small moons that are the source of the dust that forms the rings. The innermost ring, shown in gray shading, is the halo ring. The outer edge of the main ring, colored in red, coincides with the orbit of the moon Adrastea. The outer two moons Thebe and Amalthea supply the dust that forms the two disk-shaped gossamer rings, respectively (NASA/\cite{Ock1999}).}
\label{fig:jupiter2}
\end{minipage}
\hspace{0.02\columnwidth}
\begin{minipage}[b]{0.48\columnwidth}
\centering
\includegraphics[bb=0.000000 0.000000 647.280000 358.320000, width=0.8\textwidth]{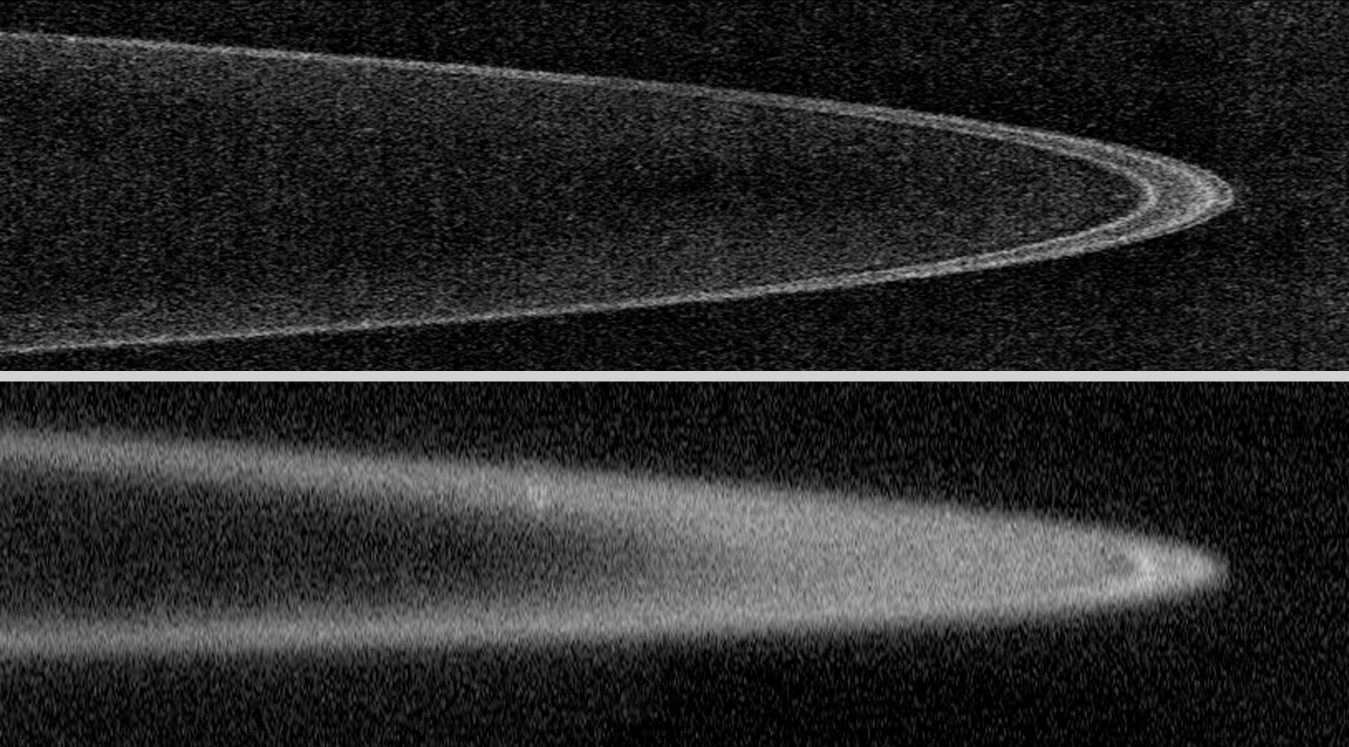}
\caption{Images of the main ring of Jupiter obtained by the New Horizons spacecraft in back-scattered (top) and forward-scattered (bottom) light. The former shows distribution of macroscopic particles, while the latter shows that of dust particles (NASA/Johns Hopkins University Applied Physics Laboratory/Southwest Research Institute).}
\label{fig:jupiter3}
\vspace{15mm}
\end{minipage}
\end{figure}

Jupiter's rings were discovered by the Voyager 1 spacecraft during its Jupiter flyby \citep{Owe1979}, and Jupiter became the third known planet that hosts rings after Saturn and Uranus.
The ring system was then investigated by Voyager 2 during its Jupiter flyby, and later in much more detail by the Galileo spacecraft, which orbited Jupiter from 1995 to 2003 \citep{Bur2004}.
Jupiter's rings consist primarily of micron-sized dust particles.
From the inside, the system is divided into the halo ring, the main ring, the Amalthea gossamer ring, and the Thebe gossamer ring (Figures \ref{fig:jupiter1} and \ref{fig:jupiter2}). 
The optical depth is very low, on the order of $10^{-6}$ even for the densest main ring, and $10^{-8}-10^{-7}$ for the two gossamer rings.
Because of these low optical depths, the ring system was not discovered by ground-based observations before the arrival of Voyager 1.

The radial extent of each ring is closely related to the orbit of the satellite that is thought to be the source of each ring (Figures \ref{fig:jupiter1} and \ref{fig:jupiter2}).
For example, the outer edge of the main ring almost coincides with Adrastea's orbit.
The outer edges of the two gossamer rings also roughly coincide with the orbits of Amalthea and Thebe, respectively, although the outer edge of the Thebe gossamer ring extends somewhat beyond Thebe's orbit.
In addition, the vertical thickness of the main ring and the two gossamer rings corresponds to the extent of the vertical excursion of the source satellite with respect to Jupiter's equatorial plane.
From these observations, it is thought that the rings were formed by the ejecta generated by meteoroid impacts with the satellite that is the source of each ring.
In the case of the main ring, many other parent bodies, in addition to Metis and Adrastea, appear to be contained in the ring and contribute to the dust production \citep{Bur2004}.
As the orbits of the ejected particles gradually shrink due to the Poynting-Robertson drag and moves radially inward, the particles become distributed in a disk shape inside the satellite orbit.
Compared to Metis and Adrastea, Amalthea and Thebe have larger orbital inclinations, so the two gossamer rings associated with the latter two satellites are thicker than the main ring.
The orbital decay timescale of dust particles due to the Poynting-Robertson drag is less than  $10^6$ years \citep{Bur2004}, so to explain the existence of Jupiter's rings, dust particles must be continuously replenished.

The core of the main ring, which is approximately 1000 km wide, contains cm-size and larger particles, while the regions inside the core is composed of dust. 
The New Horizons spacecraft on the way to Pluto observed the Jovian rings during its Jupiter flyby in 2007, and revealed the fine structure of the main ring \citep[][Figure \ref{fig:jupiter3}]{Sho2007}.

The vertical scale height of the main ring increases with decreasing distance from Jupiter. Then, at the radial location where the orbital period of dust particles and the rotation period of Jupiter (and its magnetic field) are in the 3:2 resonance, the vertical scale height of the dust increases rapidly due to the effect of electromagnetic forces in the Jovian magnetosphere, and the dust becomes distributed in a toroidal shape \citep{Bur2004}. This leads to a structure called the halo ring (Figure \ref{fig:jupiter2}).

\subsection{The rings of Neptune}\label{chap_ring:subsec34}

\begin{figure}[htb]
\centering
\begin{minipage}[b]{0.48\columnwidth}
\centering
\includegraphics[bb=0.000000 0.000000 1884.750000 1536.000000, height=0.65\textwidth]{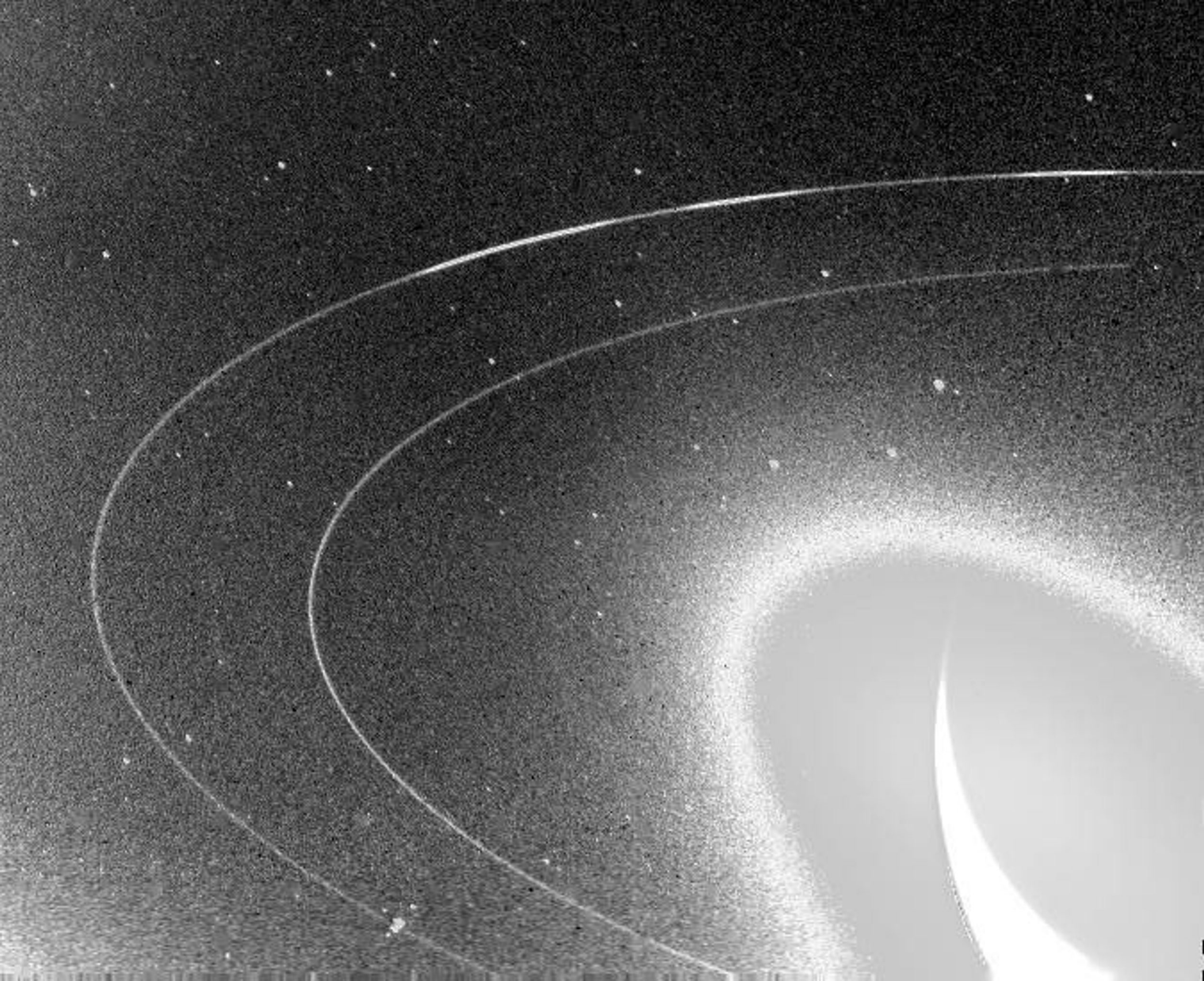}
\caption{Image of Neptune's rings taken by Voyager 2. This image was taken as Voyager 2 left the planet, when the geometry was ideal for detecting small dust particles in the rings that forward-scatter light preferentially. Bright arcs in the Adams ring can be seen (NASA/JPL).}
\label{fig:neptune1}
\end{minipage}
\hspace{0.02\columnwidth}
\begin{minipage}[b]{0.48\columnwidth}
\centering
\includegraphics[bb=0.000000 0.000000 1278.840000 807.900000, height=0.5\textwidth]{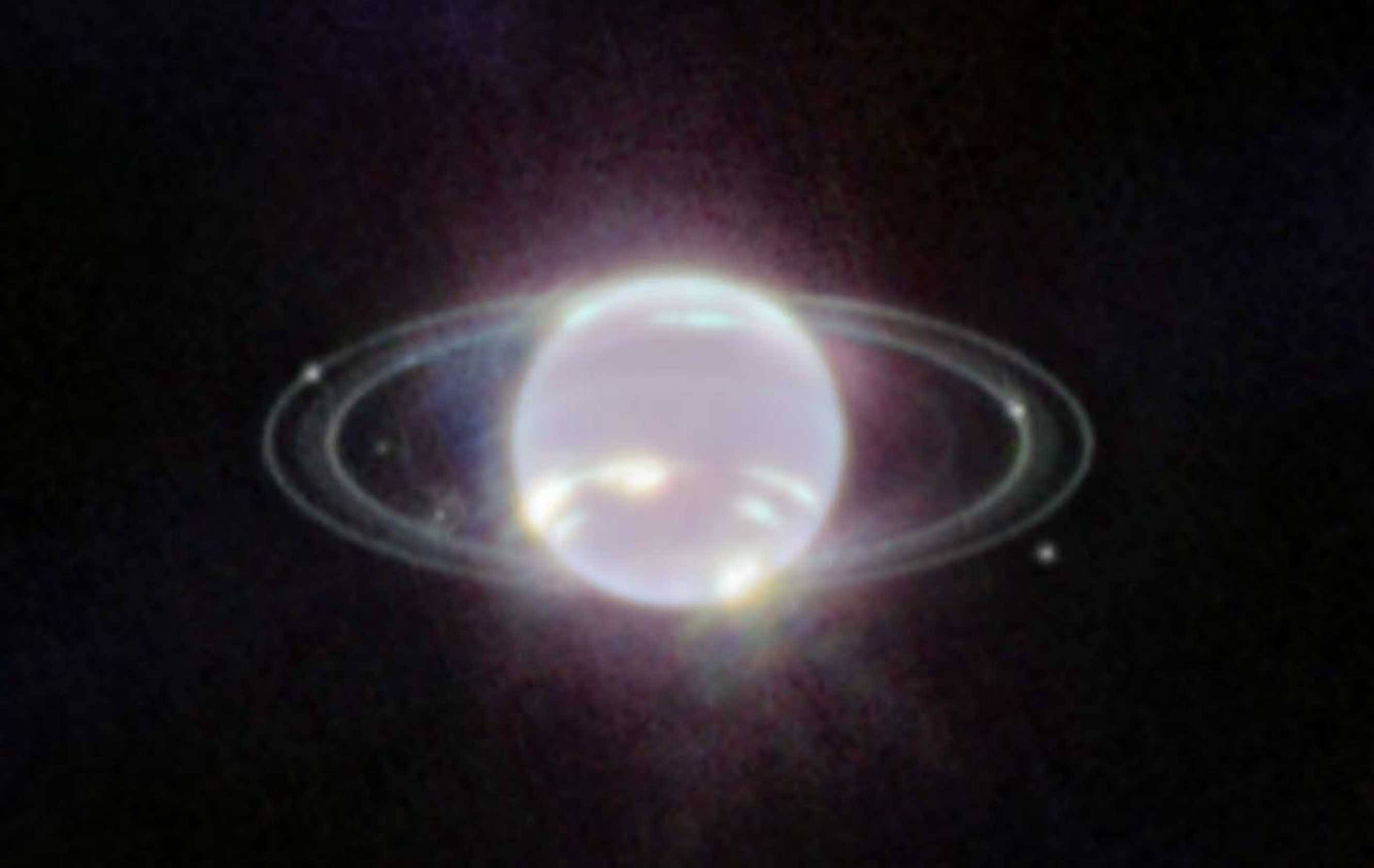}
\hspace{4mm}
\caption{Image of Neptune and its rings taken in 2022 by James Webb Space Telescope's 
Near-Infrared Camera (NASA/ESA/CSA/STScI/James Webb Space Telescope).}
\label{fig:neptune2}
\end{minipage}
\end{figure}

After the discovery of Uranus's rings, astronomers made efforts to search for Neptune's rings by also using opportunities of stellar occultation by the planet. 
Unlike the case of Uranus, during an occultation event in 1984, star light was blocked only on one side, thus, the results were interpreted as the existence of a ring-like arc \citep{Hub1986}.
Five years after the above observation, Neptune's rings were clearly confirmed by direct imaging by Voyager 2 during its Neptune flyby \citep[][Figure \ref{fig:neptune1}]{Smi1989}. 
This discovery revealed that all the four giant planets in our Solar System have rings, although they are quite different.
After the visit of Voyager 2, the Neptunian ring system was observed further with ground-based large telescopes and the Hubble Space Telescope \citep[for a review, see][]{deP2018}. 
More recently, Neptune's rings were observed by the James Webb Space Telescope (Figure \ref{fig:neptune2}).

\begin{figure}[htb]
\centering
\includegraphics[bb=0.000000 0.000000 1260.000000 1476.000000, height=0.5\textwidth]{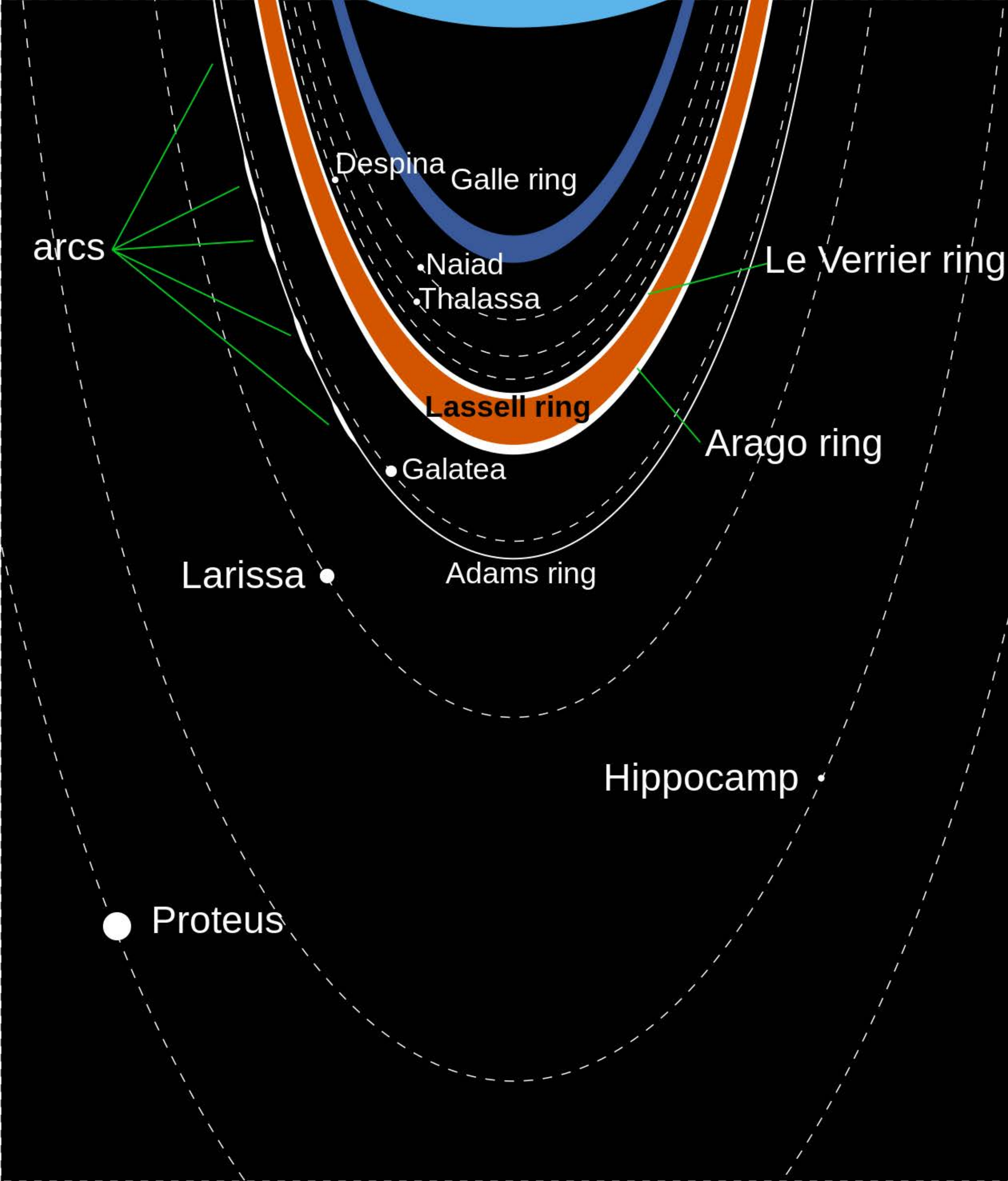}
\caption{Schematic illustration of the Neptunian ring-moon system. Solid lines denote rings, and the dashed lines denote orbits of moons. The colored bands show approximate radial extent of the dusty rings (Wikimedia).}
\label{fig:neptune3}
\end{figure}

Neptune's rings mainly consist of five rings (Figure \ref{fig:neptune3}).
From the inside, they are named Galle, Le Verrier, Lassell, Arago, and Adams.
These are the names of the astronomers who made important contributions to the discovery of Neptune and its largest moon, Triton.
Among these rings, Le Verrier, Arago, and Adams are narrow rings, and Galle and Lassell are wide rings with a low optical depth, on the order of $10^{-4}$.
The Adams ring has some azimuthal concentration of particles, called arcs, which appear bright.
It was these ring arcs that were observed during the 1984 stellar occultation.
The effect of gravity of the nearby moon Galatea likely plays an important role in the formation of these arc structures and their stability. 
From observations of Neptune's rings conducted after Voyager by large ground-based telescopes and the Hubble Space Telescope, it was confirmed that the relative intensity as well as the relative azimuthal location of the arcs in the Adams ring have changed considerably compared to the Voyager data. This suggests that the arc structures are dynamic rather than stationary.

Because the albedo of Neptune's rings is very low, the particles in the rings are likely to be covered with dark material, like the particles in the rings of Uranus.
It may be an organic material affected by radiation like the Uranus rings.
Neptune's rings have a rather high proportion of dust particles like Jupiter's rings, and are different from the rings of Saturn and Uranus.
The optical depth of the Neptunian rings is small, being less than 0.1.
As in the case of the Uranian rings, the rings of Neptune coexist with many small satellites, implying their involment in the creation of these rings.

\subsection{Rings of small solar system bodies}\label{chap_ring:subsec35}

In the Solar System, the four giant planets are not the only bodies with known rings. 
As of 2024, three other, much smaller objects named Chariklo, Haumea and Quaoar are known to contain rings, and one other object, Chiron, might also do so.

\begin{figure}[htb]
\centering
\includegraphics[bb=0.000000 0.000000 1848.000000 1256.000000, width=0.5\textwidth]{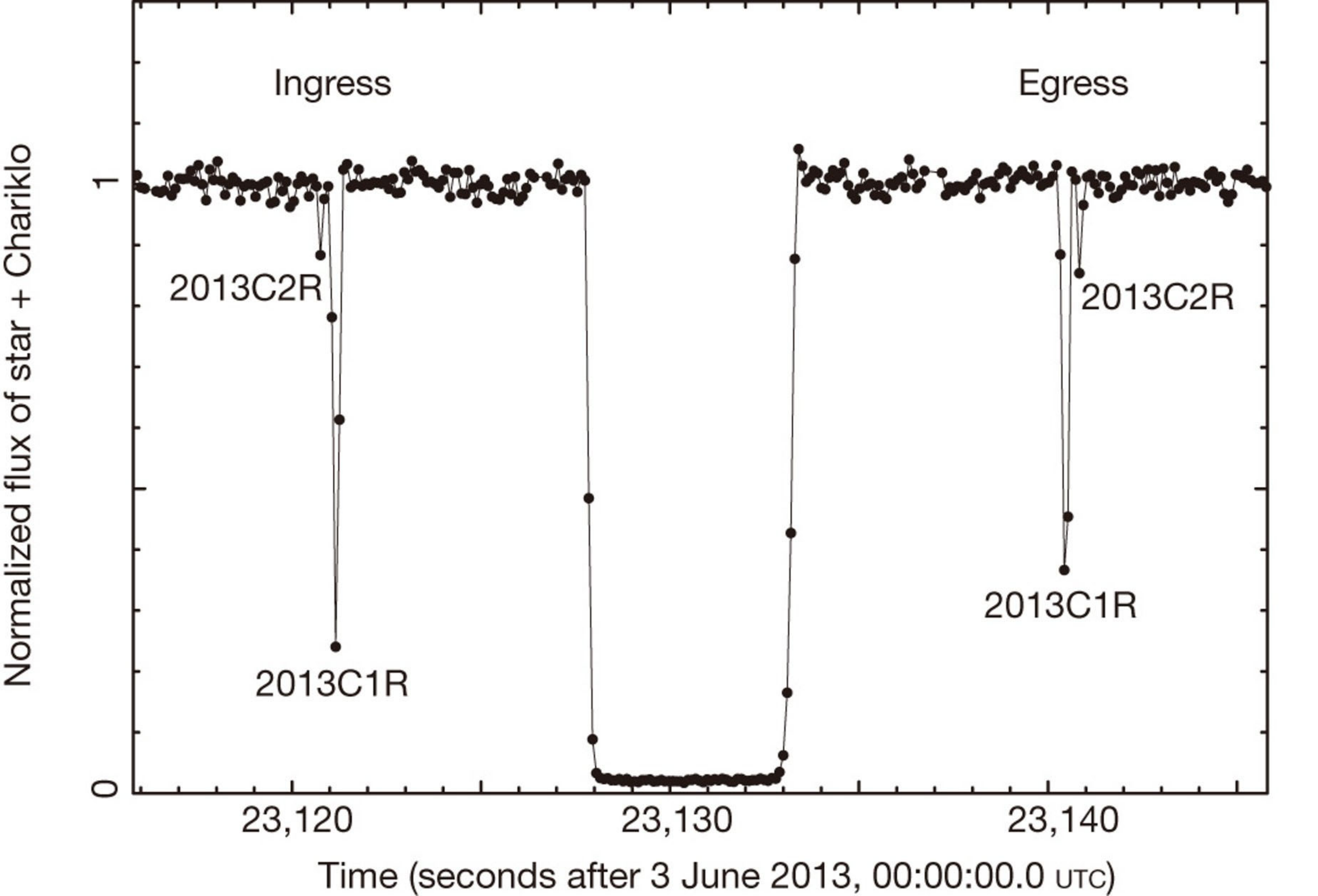}
\caption{Change of the observed flux from a star during the occultation by the Chariklo and its rings obtained by the Danish 1.5-m telescope at the European Southern Observatory at La Silla, Chile \citep{Bra2014}.
Occultations by the two rings (2013C1R and 2013C2R) at ingress (before the starlight was blocked by Chariklo) and at egress (after the star emerged from behind Chariklo) can be seen. 
}
\label{fig:chariklo}
\end{figure}

Centaurs are small Solar System bodies with orbits between Jupiter and Neptune.
Chariklo is the largest known Centaur with a radius of about 125 km, orbiting between Saturn and Uranus.
In 2013, occultation observations revealed two narrow rings around Chariklo, making it the first object with rings other than giant planets \citep[][Figure \ref{fig:chariklo}]{Bra2014}.
The two rings are located about 400 km from the center of Chariklo, with width of about 7 km and 3 km, respectively.
In fact, the presence of the rings explains characteristics of past observations of Chariklo.
Photometric observations have shown that Chariklo gradually became dimmer between 1997 and 2008, then brightened again later.
Also, reflectance spectra showed that the absorption bands due to water ice that were visible when Chariklo was bright disappeared when Chariklo was dimmer.
These can be explained by the different contributions from the rings at least partially composed of water ice with different geometries of the ring plane as seen from Earth; the orientation of the ring plane of Chariklo determined from the occultation observations shows that it was viewed almost edge-on from Earth in 2008 when it was dimmest \citep[see][for a review]{Sic2018}.

The origin of Chariklo's rings is not well understood, although several mechanisms for their origin have been proposed \citep[see, e.g.][]{Sic2018}. 
The rings may have been derived from debris generated by collisions with Chariklo, tidal disruption of a moon around Chariklo, or partial tidal disruption of Chariklo during a close encounter with a giant planet. Particles released from the interior of Chariklo due to its cometary activity may be the source of the ring material.
Collisions between the ring particles will cause the ring to radially spread, and they would migrate into Chariklo within a few million years due to the Poynting-Robertson drag.
Some mechanisms such as a resonance between the orbital period of ring particles and the rotation period of the non-spherical central body, and/or the mean motion resonance with unseen moon(s) may contribute to maintaining the narrow rings.

On the other hand, occultation observations in 2017 discovered a ring around Haumea, making it the first ringed trans-Neptunian object \citep{Ort2017}.
Haumea's radius is about 800 km, and its ring has a radius of about 2300 km with a width of about 70 km.
Haumea is a fast rotator with a rotation period of 3.9 h.
It is known to belong to a group of objects with similar orbital parameters and surface properties called ``family''.
The family members are thought to have originated in a disruptive impact of a progenitor body, which may also be related to the origin of the ring.
The orbital period of the ring is close to the 1:3 resonance with Haumea's spin period, i.e., Haumea rotates three times on its axis in the time that a ring particle completes one revolution. 
This resonance may contribute to confining the ring against radial spreading.

Recently, a ring system has been confirmed around another trans-Neptunian object, Quaoar. 
The outer ring, named Q1R, was discovered first \citep{Mor2023}, and then another ring, Q2R, was discovered further inward. 
Both rings are located at radial distances where the orbital period is an integer ratio of the rotation period of Quaoar, so spin-orbit resonances with Quaoar as well as the influence of its moon Weywot seems to play a role in maintaining their shape. 
A distinctive feature of the Quaoar ring is that the outer ring, Q1R, is located outside the Roche limit, if a particle density similar to that of Saturn's ring particles is assumed. 
Outside the Roche limit, the Hill radius of the particles becomes larger than the physical radius, thus gravitational accretion of particles is possible if sufficient energy is dissipated at collision. 
The ring may be maintained because the particles are elastic, but the details are not well understood. 

It has also been suggested that Chiron, another Centaur, may have a ring system.
However, the structure of the material surrounding Chiron appears to be changing over time, thus further observations seem to be needed to clarify the structure of the system \citep[see][]{Sic2024}.

\section{Origins of ring systems}\label{chap_ring:sec4}%

How the rings formed is closely related to the formation process of the central body that hosts the rings and its evolution after formation.
As we described in Section \ref{chap_ring:sec3}, all the four giant planets in the Solar System have rings, but their characteristics are diverse, suggesting that their origins are also diverse.
Of these, the rings of Jupiter, Uranus, and Neptune are much less massive than Saturn's rings, and are either dusty rings or dense narrow rings.
As shown in Figure \ref{fig:section2_f1} and described in Section \ref{chap_ring:sec3}, these rings are located very close to the orbits of small moons, and it is likely that these moons are related to the origin of the rings.
In contrast, it is not easy to understand the origin of the massive rings of Saturn.

One of the traditional ideas for the origin of Saturn's rings is the disruption of a small satellite by an impact of another body, such as a comet.
In order to form Saturn's rings, a satellite with a mass at least as large as Mimas needs to be disrupted, and also a sufficiently large impactor is required to destroy a satellite of that size.
Such a large impact is unlikely to occur in the current Solar System, but may be possible in the early Solar System, if a large number of trans-Neptunian objects were scattered inward, for instance, due to gravitational perturbation by giant planets that experienced orbital instability.

As an alternative mechanism to cause disruption of a satellite, a high-velocity impact between  satellites that leads to their catastrophic disruption is also proposed  \citep{Cuk2016}.
The satellites are formed outside the Roche limit $a_\mathrm{Roche}$, but if their orbits are too far away from the Roche limit, the fragments will re-accumulate to form new satellites, without delivering sufficient mass into the Roche zone to form rings \citep[see, e.g.][]{Cha2018}.
Thus, in order to supply the fragments within the Roche limit and form rings, the disruption must occur not too far from $a_\mathrm{Roche}$.
Also, if a satellite made of a mixture of rock and ice is disrupted, the rocky material has to be removed somehow to form the current icy rings.
Research on the formation of Saturn's rings by these mechanisms is still ongoing.

On the other hand, recent studies on the dynamical evolution of the rings show that the current mass of Saturn's rings can be naturally explained as an outcome of the radial spreading of primordial much more massive rings \citep{Salm2010}.
When the surface density of the ring is high enough so that self-gravity wakes are formed to lead to $Q\sim2$, gravitational interaction between the wakes plays a major role in the transport of angular momentum in the ring, and the viscosity in this case is proportional to the square of the surface density \citep{Dai2001}.
When the radial spreading of the ring is examined taking this viscosity into account, some of its mass falls into Saturn while other spreads radially outward through the Roche limit to form satellites, and it was shown that the mass remaining in the ring after 4 Gyr becomes on the order of $10^{19}$ kg regardless of the initial ring mass.
This value of the asymptotic mass is comparable to the mass of Mimas, and is on the same order as the current mass of Saturn's rings estimated from Cassini observations \citep{Ies2019}.
In other words, if Saturn's rings evolved from more massive rings than the current rings, they would evolve into rings approximately with the current mass after the age of the Solar System, regardless of their initial mass.

One possible mechanism for the formation of such a massive ring is the disruption of a Titan-sized satellite whose orbit gradually migrates toward Saturn \citep{Can2010}.
Gas giant planets such as Jupiter and Saturn are formed by accretion of gas and solids from the protoplanetary disk, and a disk of gas and solids is formed around the planet, which is called a circumplanetary disk or protosatellite disk.
Titan was likely formed by repeated collisions and merging of solids in such a disk.
As the moons grow larger, they lose angular momentum as a result of gravitational interactions with the surrounding gas in the circumplanetary disk, and migrate into the planet.
In this scenario, the rings are formed by the tidal disruption of the last large satellite that migrates into Saturn, before the circumplanetary disk has dissipated.
If a differentiated Titan-sized satellite with a rocky core and icy mantle is disrupted, the more fragile icy mantle is stripped off and the rocky core falls to Saturn, leaving debris of icy fragments
\citep{Can2010}.
The massive ring spreads radially, some of which falls to Saturn, while the particles that have spread beyond the Roche limit re-accumulate and become the current icy satellites.
Another mechanism for the formation of the massive ring is the disruption of a large trans-Neptunian object that made a close encounter with Saturn.
If a differentiated Titan-sized object is scattered into the Roche zone and its icy mantle is stripped off by the tidal force of Saturn, a massive ring of icy debris can be formed around the planet \citep{Hyo2017}.
Furthermore, an alternative model has also been proposed in which an ancient moon on a distant orbit exterior to that of Titan helped tilt Saturn's spin axis through a spin-orbit resonance between the precession of Saturn's spin and the precession of the orbit of Neptune, then this ancient moon subsequently experienced orbital instability and a grazing encounter with Saturn, leading to its tidal disruption and formation of the rings \citep{Wis2022}. In this case, the formation of the rings is estimated to have occurred around 100 million years ago to be consistent with orbital evolution.

The age of Saturn's rings provides clues to their origin, but is poorly constrained.
The rings may be polluted and darkened by meteoroid impacts since their formation.
Using the observed meteoroid flux and the present ring composition and assuming that the rings were made of pure ice initially, this model allows us to estimate the length of time the rings have been bombarded by meteoroids \citep[e.g.,][]{Kem2023}.
In this case, the fact that the current rings are almost pure ice has been interpreted as indicating that such exposure age is short, i.e., that the rings are young, and the rings may have been created by recent disruption of satellite(s) by mechanisms described above.
However, the exposure age obtained in this way requires assumptions about the meteorite flux since the time of ring formation as well as the accretion efficiency of impacting non-icy meteoroid material, and may not be necessarily the same as the age of the ring since its formation \citep{Cri2019}.
In addition, a recent study has pointed out that meteoroids impacting ring particles at high velocities evaporate and their non-icy material, when considering as very small, nanometer-sized grains, may be efficiently transported electromagnetically into Saturn's atmosphere \citep{Hsu2018}, and thus, may not be readily accreted into ring particles \citep{Hyo2025}.
In this case, the fact that Saturn's rings are mostly icy may have nothing to do with their age, and it would be even possible that the rings were formed around the time Saturn was formed.
Research on the origin and the age of Saturn's rings is still ongoing.

\section{Concluding remarks}\label{chap_ring:sec5}%

Until the mid-1970s, Saturn was the only planet with observed rings, but now it has been confirmed that all the giant planets in the Solar System have rings, and rings have also been discovered around smaller bodies.
The sizes and shapes of the rings discovered so far are diverse, which indicates that their origins and evolutionary processes should also be diverse.
We have seen that moons in the vicinity of ring regions play an important role in the formation of structures in the rings, and that the origin and evolution of each ring system seem to be closely related to the origin and evolution of nearby moons.
Thus, the origin of ring systems should be inevitably investigated as part of origin and evolution of ring-satellite systems.

Spacecraft observations revealed intriguing details about the ring systems of the giant planets.
As clearly demonstrated by the examples of Jupiter and Saturn, observations from a spacecraft in orbit about the planet significantly advanced our understanding of the ring system, providing numerous new discoveries and a wide range of new data at a much higher resolution than flyby observations, covering longer periods and making it possible to study seasonal changes. 
For this reason, exploration of the ring-satellite system of Uranus by an orbiter is highly desirable.
Sending an orbiter to Neptune is also desirable for more detailed investigation of its ring system, although it is more challenging because of the distance from Earth and the fact that it has only a single large satellite, Triton, in a retrograde orbit relatively distant from the planet, which makes the orbital design for ring observations more difficult.
As for the rings of Saturn, if future spacecraft mission allows close-proximity observations of individual ring particles and their impact processes, it will not only help us understand the physical properties of ring particles, but also greatly advance our understanding of the origin and evolution of the ring-satellite systems \citep{Spi2018}.

If further occultation observations found ring systems around many more small bodies to allow us to conduct statistical analysis, such as the relationship between the probability of finding ring systems and surface properties of the ring-hosting small bodies, our understanding of the origins of their ring systems as well as the dynamical history of the host small bodies would significantly improve.
Spacecraft exploration of ringed small bodies will also be needed to clarify their fine structures,  to understand dynamical and physical relationships between rings and nearby moons, to find currently undiscovered moons, and to clarify the relationship between the cometary activity of the central body and the formation of its rings.

Methods to search for rings around exoplanets have been proposed, and some constraints have been derived through data analysis \citep[e.g.,][]{Aiz2018}.
While unusual transit data have been interpreted as giant planetary rings with a radius of 0.6 au \citep{Ken2015} and possibility of some of ``super-puffs'', a class of exoplanets with exceptionally large radii, being ringed planets has been suggested \citep[e.g.,][]{Aki2020}, ring systems like those of Saturn have yet to be discovered.
On the other hand, detection of a circumplanetary disk composed of gas and dust around a giant planet PDS70c has been reported \citep[e.g.,][]{Ben2021}.
Large moons in nearly circular prograde orbits such as the Galilean moons and Titan are thought to have formed in circumplanetary disks, and their formation processes may be related to the origin of rings as we discussed in Section \ref{chap_ring:sec4}.
Not only detection of already-formed ring systems around exoplanets but also observation of circumplanetary disks that are likely to be the birth place of ring-moon systems will be important for a better understanding of the formation and evolution of rings around exoplanets.
Just as research on exoplanets and protoplanetary disks have greatly advanced our understanding of the origin of our Solar System, it is expected that observation and theoretical research on exoplanetary ring systems and circumplanetary disks around exoplanets will provide new insights into the origin and evolution of ring-satellite systems of planets and small bodies in our Solar System.

\begin{ack}[Acknowledgments]
 
I am deeply grateful to Dimitri Veras, J{\"u}rgen Schmidt, Heikki Salo, Glen Stewart, and Ren Ikeya for their thoughtful comments and useful suggestions to improve the presentation of the manuscript.
I would also like to thank Douglas Hamilton for providing the original file for Figure \ref{fig:section2_f1}, and Takaaki Takeda for his prompt and invaluable assitance in the preparation of Figures \ref{fig:section2_f1} and \ref{fig:saturn1}.
I acknowledge support from JSPS KAKENHI Grant No. JP23K22557.

For further reading, chapters in the following books present comprehensive reviews on the state-of-the-art knowledge at the time of publication of the Saturnian system including its rings and of various topics on planetary ring systems, respectively: \citet{Dou2009}, \citet{Tis2018}.
Each of the following textbooks contains an excellent chapter on planetary rings including detailed explanations for various basic processes: \citet{Mur1999}, \citet{deP2015}.
Also, there are many excellent reviews on ring systems from various points of view, such as
\citet{Tis2013}, \citet{Cha2018b}, and \citet{Sic2024}.

\end{ack}

\bibliographystyle{Harvard}
\bibliography{reference}

\end{document}